\documentclass[twocolumn,showpacs,amsmath,amssymb,aps,prd,floats]{revtex4-1}
\usepackage{graphicx}
\usepackage[switch,pagewise,running]{lineno}
\usepackage{dcolumn}
\usepackage{bm}
\usepackage{epsfig}
\usepackage{xspace}
\usepackage{hyperref}
\hypersetup{colorlinks,linkcolor=red,citecolor=blue,urlcolor=blue}  
\usepackage{color,xcolor}
\usepackage{verbatim}
\usepackage{float}
\usepackage{slashed}
\usepackage{rotating}
\usepackage[normalem]{ulem} 
\usepackage[caption=false]{subfig}

\begin{document}

\title{Hadronuclear interactions in the jet of low TeV luminosity AGN: Implications for the low-state very-high-energy gamma-ray emission}
\author{Rui Xue$^{1}$}
\author{Ze-Rui Wang$^{2}$}\email{zerui\_wang62@163.com}
\author{Wei-Jian Li$^{1}$}

\affiliation{$^1$Department of Physics, Zhejiang Normal University, Jinhua 321004, China\\ 
$^2$College of Physics and Electronic Engineering, Qilu Normal University, Jinan 250200, China}

\begin{abstract}
In the study of radiation mechanisms of AGNs' jets, hadronuclear ($pp$) interactions are commonly neglected, because the number density of cold protons in the jet is considered insufficient. Very recently, our previous work proves that $pp$ interactions in the low TeV luminosity AGNs, which have potential to generate detectable very-high-energy (VHE) emission, could be important. Based on this, the one-zone $pp$ model is employed to study low-state quasi-simultaneous spectral energy distributions of a sample of low TeV luminosity AGNs in this work. Our modeling results show that the $\gamma$-ray generated in $pp$ interactions can explain the observed TeV spectra and has contribution to higher energy band that could be detected by the Large High Altitude Air Shower Observatory (LHAASO). In the sample, we suggest that M 87, Mrk 421 and Mrk 501 are the most likely objects to be detected by LHAASO in the near future. Other possible origins of VHE emission are also briefly discussed.
\end{abstract}

\maketitle

\section{Introduction}
Being some of the most powerful persistent objects of electromagnetic radiation in the universe, the jet-dominated active galactic nuclei (AGNs), generally including blazars and radio galaxies (RGs), dominate the diffuse extragalactic $\gamma$-ray background \citep{2015ApJ...800L..27A, 2016PhRvL.116o1105A}. In the unification model of AGNs \citep{1995PASP..107..803U}, blazars are the subclass with relativistic jets pointing to observers. Based on their optical spectra, blazars are divided into BL Lacertae objects (BL Lacs) with weak or no emission lines  ($\rm EW$ $<$ $\rm 5\AA$) and flat spectrum radio quasars with stronger emission lines ($\rm EW \ge 5\AA$). In contrast to blazars, RGs are viewed at a substantial inclination to the jet axis. According to the jet luminosity at 178 MHz, RGs are classified as Fanaroff--Riley (FR) I ($< 5\times 10^{24}~\rm W~Hz^{-1}~Sr^{-1}$) and II ($\geqslant 5\times 10^{24}~\rm W~Hz^{-1}~Sr^{-1}$) sources \citep{1974MNRAS.167P..31F}. Generally, FR I RGs are believed to be the parent population of BL Lacs. In the TeVcat catalog \footnote{http://tevcat.uchicago.edu/}, most TeV blazars are BL Lacs and all the TeV RGs are FR I RGs. 

The broadband spectral energy distributions (SEDs) of jet-dominated AGNs display two bumps. Due to the detection of significant linear polarization \citep{1997A&A...325..109S}, it is generally accepted that the low-energy bump, from radio to optical/X-ray, originates from the synchrotron radiation of primary relativistic electrons that accelerated in the jet. According to peak frequency $\nu_{\rm S,peak}$ of the low-energy bump, blazars are divided into low synchrotron peaked (LSP; $\nu_{\rm S,peak}\lesssim 10^{14}~\rm Hz$), intermediate synchrotron peaked (ISP; $10^{14}~\rm Hz \lesssim \nu_{\rm S,peak}\lesssim 10^{15}~\rm Hz$), and high synchrotron peaked (HSP; $\nu_{\rm S,peak}\gtrsim 10^{15}~\rm Hz$) sources \citep{2010ApJ...716...30A, 2016ApJS..226...20F, 2022ApJS..262...18Y}. At present, the origin of the high-energy bump, from X-ray to $\gamma$-ray, is debated. In leptonic scenarios, the high-energy bump is explained by the inverse Compton (IC) radiation of primary relativistic electrons that up-scatter soft photons, which could be dominated by the synchrotron photons emitted by the same population of electrons within the jet (synchrotron self-Compton, SSC; e.g., \citep{1981ApJ...243..700K}), or by the photons from the fields outside the jet (external inverse Compton, EC; e.g., \citep{1994ApJ...421..153S}). In hadronic scenarios, both the proton synchrotron radiation \citep{2000NewA....5..377A, 2013ApJ...768...54B, 2015MNRAS.448..910C} and the emission from secondary particles \citep{2017MNRAS.464.2213P} that generated in the hadronic interactions and internal $\gamma \gamma$ pair production account for the high-energy bump. 

When modeling SEDs of jet-dominated AGNs, the most commonly applied model is the one-zone model, which assumes that all the jet's non-thermal emission comes from a compact spherical region (e.g., \citep{2009MNRAS.397..985G, 2021MNRAS.506.5764D}). Since various soft photon fields (such as the jet inside, the sheath structure of jet, the accretion disc, the broad-line region and the dusty torus) exist in the AGNs environment, photohadronic ($p\gamma$) interactions, including the photopion production and Bethe-Heitler pair production, are naturally considered as the most likely hadronic processes. However, $p\gamma$ interactions seems to fail on interpreting some observational phenomena in the framework of the one-zone model. For example, the one-zone $p\gamma$ model can only predict a very low annual neutrino detection rate for recently discovered association events between blazars and high-energy neutrinos \citep{2019MNRAS.483L..12C, 2019NatAs...3...88G, 2019ApJ...886...23X, 2020ApJ...899..113P, 2020ApJ...891..115P, 2021JCAP...10..082O}. In addition, when fitting the hard-TeV spectra of blazars, since the $p\gamma$ interactions are very inefficient, the required jet power would be 6--7 magnitude higher than the Eddington luminosity of the central super-massive black hole (SMBH) \citep{2014ApJ...783..108C, 2022ApJ...925L..19C}, even comparable to the characteristic energy of a $\gamma$-ray burst \citep{2018pgrb.book.....Z}.

In the study of investigating the radiation mechanisms of AGNs' jets, the hadronuclear ($pp$) interactions for blazars are commonly neglected, since the particle density in the jet is believed insufficient \citep{2003ApJ...586...79A}, and for RGs, $pp$ interactions are usually considered in large-scale components of jet, such as giant lobes \citep{2012ApJ...753...40F, 2013ApJ...770L...6S, 2016A&A...595A..29S, 2021MNRAS.500.1087B}. Recently, our previous work \citep{2022A&A...659A.184L} provides an analytical method to study if $pp$ interactions could be important for blazars. Analytical results suggest that sufficient cold protons may exist in the jet of low TeV luminosity AGNs, so that $pp$ interactions can contribute to the TeV spectrum without introducing extreme physical parameters. Moreover, since the $\gamma$-ray spectrum produced from $\pi_0$ decay in the $pp$ interactions basically follows the spectrum of primary protons, it may has a wide spectrum in very-high-energy (VHE, $0.1\rm~TeV\sim 100~TeV$) $\gamma$-ray band, which could be detected by the Large High Altitude Air Shower Observatory (LHAASO) \citep{2019arXiv190502773C} in the near future. In this work, following our previous analytical results \citep{2022A&A...659A.184L}, we collect a sample of low TeV luminosity AGNs, and revisit their quasi-simultaneous multi-wavelength SEDs in the framework of the one-zone $pp$ model, and study if the emission from $pp$ interactions in the jet has potential contribution to the VHE band. This paper is organized as follows. The description of the one-zone $pp$ model is presented in Section~\ref{model}. In Section~\ref{app}, we apply our model to a sample of 12 low TeV luminosity AGNs.  Finally, we end with discussion and conclusion in Section~\ref{DC}. The cosmological parameters $H_{0}=69.6\ \rm km\ s^{-1}Mpc^{-1}$, $\Omega_{0}=0.29$, and $\Omega_{\Lambda}$= 0.71 \citep{2014ApJ...794..135B} are adopted in this work.

\begin{figure}[htbp]
\includegraphics[width=0.48\textwidth]{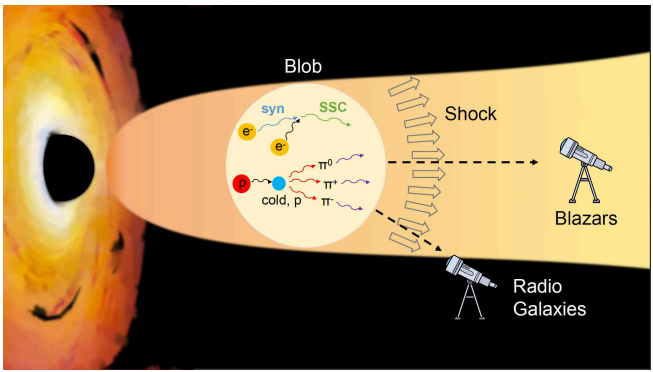}
\caption{A schematic illustration (not to scale) of the one-zone $pp$ model. It is assumed that all the jet's emission is from a single spherical region, also known as the blob. In the blob, accelerated relativistic electrons (the yellow circles) emit synchrotron and SSC emissions, and accelerated relativistic protons (the red circle) interact with cold protons (the blue circle) that come from the jet inside.  
\label{sketch}}
\end{figure}

\section{Model description}\label{model}
Based on the conventional one-zone leptonic model, we further consider hadronic emission from $pp$ interactions, in which cold protons are from the jet inside. Currently, the jet composition is uncertain \citep{2021ApJ...906..105C}. Both a normal jet with pure $e^--p$ plasma and a jet with pure $e^{\pm}$ pair plasma have been extensively studied \citep{2014Natur.515..376G, 2016ApJ...831..142M, 2020ApJS..248...27T, 2021A&A...654A..96Z, 2022ApJ...928L...9Z}. In our model, we do not specify the jet composition, but assume that the total jet power is dominated by relativistic and cold protons \citep{2022A&A...659A.184L}, where cold protons are assumed to be the part of protons that are not accelerated in the jet. A sketch of our model is shown in Figure~\ref{sketch}. In the framework of one-zone model, it is assumed that all the jet's non-thermal radiation is from a single spherical region (hereafter referred to as the blob) composed of a plasma of charged particles in a uniformly entangled magnetic field $B$ with radius $R$ and moving with the bulk Lorentz factor $\Gamma=\frac{1}{\sqrt{1-\beta_{\Gamma}^2}}$, where $\beta_{\Gamma} c$ is the speed of the blob, at a viewing angle $\theta_{\rm obs}$ with respect to the line of sight of the observer. For blazars, since the jet is pointing to us, the observed emission will be Doppler boosted by a factor $\delta_{\rm D}^4$, where $\delta_{\rm D}=[\Gamma(1-\beta_{\Gamma} \rm cos\theta_{\rm obs})]^{-1}$ is the Doppler factor. It is usually difficult to constrain $\theta_{\rm obs}$, therefore we simply set $\delta_{\rm D} \approx \Gamma$ by assuming $\theta_{\rm obs} \lesssim 1/\Gamma$ in the modeling of blazars' emission. For RGs, $\delta_{\rm D}$ is determined by introducing a reasonable value of $\theta_{\rm obs}$ as suggested in observations. In the following, all quantities are measured in the comoving frame, unless specified otherwise.

\subsection{Particle energy distribution}\label{ped}
Relativistic electrons are assumed to be injected with a smooth broken power-law energy distribution at a constant rate given by
\begin{equation}
\begin{split}
    \dot{Q}^{\rm inj}_{\rm e}(\gamma_{\rm e})=\dot{Q}_{\rm e,0}\gamma_{\rm e}^{-\alpha_{\rm e,1}}[1+(\frac{\gamma_{\rm e}}{\gamma_{\rm e,b}})^{(\alpha_{\rm e,2}-\alpha_{\rm e,1})}]^{-1}, \\
    \gamma_{\rm e,min}<\gamma_{\rm e}<\gamma_{\rm e,max},
\end{split}
\end{equation}
where $\dot{Q}_{\rm e,0}$ is the normalization in units of $\rm cm^{-3}~s^{-1}$, $\gamma_{\rm e,min/b/max}$ are the minimum, break, and maximum electron Lorentz factors, $\alpha_{\rm e,1}$ and $\alpha_{\rm e,2}$ are the electron spectral indices before and after $\gamma_{\rm e,b}$. By giving an electron injection luminosity $L_{\rm e,inj}$, $\dot{Q}_{\rm e,0}$ can be calculated by $\int \dot{Q}^{\rm inj}_{\rm e}\gamma_{\rm e}m_{\rm e}c^2 d\gamma_{\rm e}=\frac{3}{4\pi R^3}L_{\rm e,inj}$, where $m_{\rm e}$ is the electron rest mass, $c$ is the speed of light. When the radiative cooling and/or particle escape balances with injection, a steady-state electron energy distribution (EED) is achieved, which can be written as
\begin{equation}
   N_{\rm e}(\gamma_{\rm e})\approx \dot{Q}^{\rm inj}_{\rm e}(\gamma_{\rm e})t_{\rm e},
\end{equation}
where $t_{\rm e}={\rm min}\{t_{\rm cool}, t_{\rm dyn}\}$. More specifically, $t_{\rm dyn}=R/c$ is the dynamical timescale of the blob, and $t_{\rm cool}=\frac{3m_{\rm e}c}{4\gamma_{\rm e}\sigma_{\rm T}(U_{\rm B}+\kappa_{\rm KN}U_{\rm syn})}$ is the radiative cooling timescale, where $\sigma_{\rm T}$ is the Thomson scattering cross section, $U_{\rm B}=B^2/8\pi$ is the energy density of the magnetic field, $U_{\rm syn}$ is the energy density of synchrotron photons, and 
\begin{equation}
\begin{split}
\kappa_{\rm KN}=\frac{9}{U_{\rm syn}}\int^{\infty}_0dE~E~n_{\rm syn}(E) \\ 
\int^1_0 dq\frac{2q^2\textmd{ln}~q+q(1+2q)(1-q)+\frac{q(\omega q)^2(1-q)}{2(1+\omega q)}}{(1+\omega q)^3}
\end{split}
\end{equation}
is a numerical factor accounting for the Klein-Nishina effect \citep{2010NJPh...12c3044S}, where $E$ is the energy of target photons, $n_{\rm syn}(E)$ is the number density distribution of synchrotron photons, $\omega=4E \gamma_{\rm e}/(m_{\rm e}c^2)$. In consideration of the EED evolution, SSC process causes nonlinear cooling effects \citep{2017AIPC.1792e0022Z}. In our modeling, based on the assumption that the observed SEDs of our sample are dominated by the emission from primary electrons, by introducing the Compton parameter $Y$, which represents the ratio of fluxes or luminosities of high-energy and low-energy bumps, $t_{\rm cool}$ can be estimated as $t_{\rm cool}=\frac{3m_{\rm e}c}{4\gamma_{\rm e}\sigma_{\rm T}U_{\rm B}(1+Y)}$. 

Relativistic protons are assumed to be injected with a power-law energy distribution at a constant rate:
\begin{equation}
    \dot{Q}^{\rm inj}_{\rm p}(\gamma_{\rm p})=\dot{Q}_{\rm p,0}\gamma_{\rm p}^{-\alpha_{\rm p}}, \gamma_{\rm p,min}<\gamma_{\rm p}<\gamma_{\rm p,max},
\end{equation}
where $\dot{Q}_{\rm p,0}$ is the normalization in units of $\rm cm^{-3}~s^{-1}$, $\gamma_{\rm p,min/max}$ are the minimum, and maximum proton Lorentz factors, $\alpha_{\rm p}$ is the spectral index. Similarly, by giving a proton injection luminosity $L_{\rm p,inj}$, $\dot{Q}_{\rm p,0}$ can be determined by $\int \dot{Q}^{\rm inj}_{\rm p}\gamma_{\rm p}m_{\rm p}c^2 d\gamma_{\rm p}=\frac{3}{4\pi R^3}L_{\rm p,inj}$, where $m_{\rm p}$ is the proton rest mass. Then the steady-state proton energy distribution (PED) can be written as
\begin{equation}
    N_{\rm p}(\gamma_{\rm p})\approx \dot{Q}^{\rm inj}_{\rm p}(\gamma_{\rm p})t_{\rm p},
\end{equation}
where $t_{\rm p}={\rm min}\{t_{\rm pp},t_{\rm dyn},t_{\rm p,syn}\}$ with $t_{\rm p,syn}=\frac{6\pi m_{\rm e}c^2}{(c\sigma_{\rm T}B^2\gamma_{\rm p})(m_{\rm p}/m_{\rm e})^3}$ being the proton-synchrotron cooling timescale, and $t_{\rm pp}=\frac{1}{K_{\rm pp}\sigma_{\rm pp}n_{\rm H}c}$ being the cooling timescale of $pp$ interactions, where $n_{\rm H}$ represents the number density of cold protons in the jet, $K_{\rm pp}\approx 0.5$ is the inelasticity coefficient, and $\sigma_{\rm pp}$ is the cross section for inelastic $pp$ interactions \citep{2006PhRvD..74c4018K}
\begin{equation}
\sigma_{\rm pp} = (34.3+1.88L+0.25L^2)\left[1-(\frac{E_{\rm th}^{\rm pp}}{E_{\rm p}})^4\right]^2,
\end{equation}
where $E_{\rm p}=\gamma_{\rm p}m_{\rm p}c^2$ is the relativistic proton energy, $E_{\rm th}^{\rm pp}=1.22\times10^{-3}~\rm TeV$ is the threshold energy of production of $\pi^0$, and $L=\rm{ln}(\it{E}_{\rm p}/\rm 1~TeV)$.

\subsection{Leptonic emission}
After obtaining the steady-state EED $N_{\rm e}(\gamma_{\rm e})$, we adopt the public Python package \texttt{NAIMA} \footnote{https://naima.readthedocs.io/en/latest/} to calculate the synchrotron, and IC emissions from blob and to correct the GeV-TeV spectrum absorbed by the extragalactic background light (EBL) \citep{2015ICRC...34..922Z}. 

In the modeling of synchrotron emission, we further calculate the synchrotron self-absorption (SSA), which is not considered in \texttt{NAIMA} but is very important in AGNs' jet. The SSA coefficient is obtained by \citep{1979rpa..book.....R}
\begin{equation}
    k_{\rm syn}(\nu) = -\frac{1}{8\pi \nu^2 m_{\rm e}} \int d\gamma_{\rm e} P(\nu, \gamma_{\rm e})\gamma^2_{\rm e} \frac{\partial}{\partial \gamma_{\rm e}}[\frac{N(\gamma_{\rm e})}{\gamma_{\rm e}^2}],
\end{equation}
where $\nu$ is the photon frequency, $P(\nu, \gamma_{\rm e})$ is the synchrotron emission coefficient for a single electron integrated over the isotropic distribution of pitch angles. Then we can calculate the synchrotron intensity using the radiative transfer equation \citep{1999ApJ...514..138K}
\begin{equation}
\begin{split}
    I_{\rm syn}(\nu) = \frac{j_{\rm syn}(\nu)}{k_{\rm syn}(\nu)}[1-\frac{2}{\tau_{\rm SSA}(\nu)^2}(1-\tau_{\rm SSA}(\nu) e^{-\tau_{\rm SSA}(\nu)} \\ 
    -e^{-\tau_{\rm SSA}(\nu)})],
\end{split}
\end{equation}
where $j_{\rm syn}(\nu)$ is the synchrotron emission coefficient calculated by \texttt{NAIMA}, and $\tau_{\rm SSA}(\nu)= 2Rk_{\rm syn}(\nu)$ is the SSA optical depth.

Strong emission lines are not detected in our sample (see Table~\ref{table1}), which indicates that there are no strong external photon fields, therefore we neglect the photons from external fields and only calculate SSC emission in the modeling.

\subsection{Hadronic emission}
The $pp$ interactions generate neutral and charged pions, which are short-lived and eventually decay into secondary particles, including $\gamma$-ray photons, electrons/positrons, and neutrinos, i.e.,
\begin{eqnarray*}
p+p&\to& \pi^0\to \gamma+\gamma\\
p+p&\to& \pi^+\to \nu_\mu+\mu^+\to \nu_\mu+e^++\nu_{\rm e}+\bar{\nu}_\mu\\
p+p&\to& \pi^-\to \bar{\nu}_\mu+\mu^-\to \bar{\nu}_\mu+e^-+\bar{\nu}_{\rm e}+\nu_\mu.
\end{eqnarray*}
The $pp$ interactions efficiency $f_{\rm pp}$ can be estimated through
\begin{equation}
f_{\rm pp}=K_{\rm pp}\sigma_{\rm pp}n_{\rm H}R.
\end{equation}
Here we consider the total jet power $L_{\rm jet}$ as a fraction of the Eddington luminosity of the SMBH $L_{\rm Edd}$, which is usually seen as the upper limit of the jet power \citep{2015MNRAS.450L..21Z, 2022ApJ...927...33S}, i.e., $L_{\rm jet}=\xi L_{\rm Edd}, \xi \leqslant 1$. As the power of the relativistic and nonrelativistic electrons is normally negligible compared to that of relativistic and nonrelativistic protons, the number density of cold protons in the jet $n_{\rm H}$ can be estimated by \citep{2022A&A...659A.184L}
\begin{equation}\label{n_H}
    n_{\rm H}=\frac{(1-\chi_{\rm p})\xi L_{\rm Edd}}{\pi R^2\Gamma^2m_{\rm p}c^3},
\end{equation}
where $\chi_{\rm p}$ represents the ratio of the proton injection power to the total jet power. 

\begin{table*}
\renewcommand\arraystretch{1.2}
\setlength\tabcolsep{5pt}
\caption{The Sample. Columns from left to right: (1) the source name. (2) right ascension (R.A.). (3) declination (Decl.). (4) the redshift of the source. (5) the SMBH mass in units of the solar mass, $M_{\odot}$. (6) references that provide the (quasi-)simultaneous SEDs. (7) the type of jet-dominated AGNs. For TXS 0210+515, 1ES 2037+521, RGB J0152+017, 1ES 1741+196,  and RGB J2042+244, in absence of an estimated black hole mass, we considered an average value of $10^9 M_\odot$ \citep{2017ApJ...851...33P, 2022ApJ...925...40X}.}\label{tabel1}
\centering
\begin{tabular}{ccccccc}
\hline\hline
Source name	&	R.A. (J2000) 	&	Decl. (J2000)	&	$z$	&	$M_{\rm BH}$	&	SED Ref.	&	Type	\\
~(1) & (2) & (3) & (4) & (5) & (6) & (7) \\
\hline
M 87	&	12 30 47.2	&	+12 23 51	&	0.0044	&	$6.5\times10^9$ \citep{2019ApJ...875L...6E}	&	Ref.~\citep{2015MNRAS.450.4333D, 2020MNRAS.492.5354M, 2021ApJ...911L..11E}	&	FR I RGs	\\
IC 310	&	03 16 43.0	&	+41 19 29	&	0.0189	&	$1\times10^9$	\citep{2006AA...456..439K, 2017AA...603A..25A} &	Ref.~\citep{2017AA...603A..25A}	&	FR I RGs	\\
3C 264	&	11 45 05.0	&	+19 36 23	&	0.021718	&	$5\times10^8$ \citep{2015AA...581A..33D}	&	Ref.~\citep{2020ApJ...896...41A}	&	FR I RGs	\\
Mrk 421	&	11 04 19	&	+38 11 41	&	0.031	&	$1.3\times10^9$ \citep{2002AA...389..742W}	&	Ref.~\citep{2011ApJ...736..131A}	&	BL Lac-HSP	\\
Mrk 501	&	16 53 52.2	&	+39 45 37	&	0.034	&	$4.17\times10^8$ \citep{2012ApJ...759..114C}	&	Ref.~\citep{2011ApJ...727..129A}	&	BL Lac-HSP	\\
1ES 2344+514	&	23 47 04	&	+51 42 49	&	0.044	&	$6.31\times10^8$ \citep{2003ApJ...583..134B}	&	Ref.~\citep{2013AA...556A..67A}	&	BL Lac-HSP	\\
TXS 0210+515	&	02 14 17.9	&	+51 44 52	&	0.049	&	$1\times10^9$	&	Ref.~\citep{2020ApJS..247...16A}	&	BL Lac-HSP	\\
1ES 2037+521	&	20 39 23.5	&	+52 19 50	&	0.053	&	$1\times10^9$	&	Ref.~\citep{2020ApJS..247...16A}	&	BL Lac-HSP	\\
RGB J0152+017	&	01 52 33.5	&	+01 46 40.3	&	0.08	&	$1\times10^9$	&	Ref.~\citep{2008AA...481L.103A}	&	BL Lac-HSP	\\
1ES 1741+196	&	17 44 01.2	&	+19 32 47	&	0.084	&	$1\times10^9$	&	Ref.~\citep{2017MNRAS.468.1534A}	&	BL Lac-HSP	\\
RGB J2042+244	&	20 42 06	&	+24 26 52.3	&	0.104	&	$1\times10^9$	&	Ref.~\citep{2020ApJS..247...16A}	&	BL Lac-HSP	\\
1ES 0229+200	&	02 32 53.2	&	+20 16 21	&	0.14	&	$1.45\times10^9$	\citep{2012AA...542A..59M} &	Ref.~\citep{2014ApJ...782...13A}	&	BL Lac-HSP	\\
\hline
\label{table1}
\end{tabular}
\end{table*}

With the steady-state PED $N_{\rm p}(\gamma_{\rm p})$ obtained in Section~\ref{ped}, the differential spectrum of decayed $\gamma$-ray photons, electrons/positrons and neutrinos can be calculated with analytical expressions that developed by Ref. \citep{2006PhRvD..74c4018K}. The energy distribution of $\gamma$-ray induced pair cascades is evaluated using a semianalytical method that developed by Ref.~\cite{2013ApJ...768...54B}. In fact, the internal $\gamma \gamma$ opacity contributed by the synchrotron photons from primary electrons is smaller than unity (the internal $\gamma \gamma$ opacity of each object in our sample is given in Appendix~\ref{gg}), so the secondary electrons/positrons are mainly contributed by the $pp$ interactions. The magnetic field, soft photons from the primary and secondary electrons/positrons are all included as targets when considering the cooling of secondary electrons/positrons. The absorbed $\gamma$-ray photons will be redistributed at lower energies through synchrotron and SSC emissions from pair cascades. In AGNs' environment, cascades emission normally contribute from hard X-ray to higher energy band \citep[e.g.,][]{2019MNRAS.483L..12C}, which would not significantly enhance the cooling of primary electrons because of the severe KN effect.

After obtaining intensities of leptonic $I_{\rm lep}$ and hadronic $I_{\rm had}$ processes, the observed flux density can be calculated by
\begin{equation}
    F_{\rm obs}(\nu_{\rm obs}) = \frac{\pi R^2 \delta_{\rm D}^{3}(1+z)}{D_{\rm L} ^2} (I_{\rm lep}+ I_{\rm had})e^{-\tau_{\gamma\gamma}^{\rm EBL}},
\end{equation}
where $D_{\rm L}$ is the luminosity distance \citep{2009ApJ...703.1939V}, $z$ is the redshift, ${\nu}_{\rm obs}$ = $\nu \delta_{\rm D}/(1+z)$, and $\tau_{\gamma\gamma}^{\rm EBL}$ is the optical depth for the EBL absorption \citep{2011MNRAS.410.2556D}. The SED of EBL shows two main components, one of which is an optical component with peak around 1~eV contributed by starlight and the other is an infrared component with peak around $1\times10^{-2}\rm~eV$ originated from reprocessing of starlight by dust \citep{2017A&A...606A..59H}. Depending on the redshift of object, the emitted photons with energy from $\gtrsim 2m_{\rm e}^2c^4/\rm 1~eV\approx 0.52~TeV$ to $\sim 2m_{\rm e}^2c^4/\rm 0.01~eV\approx 52~TeV$ will be absorbed due to interactions with EBL. In our modeling, since the $\gamma$-ray spectrum produced by $\pi^0$ decay in $pp$ interactions basically follows the spectrum of primary PED, this $\gamma$-ray spectrum may even extend to ultra-high-energy (UHE, $\geqslant100\rm~TeV$) $\gamma$-ray band. However, UHE photons can hardly be observed from AGNs since they will be absorbed by the cosmic microwave background (CMB) during the propagation (e.g., see Fig. 5 of Ref. \citep{2018APh...102...39H}).

\section{Application}\label{app}
As indicated by Ref.~\citep{2022A&A...659A.184L}, if the $\gamma$-ray generated in the $pp$ interactions has contribution to the TeV spectrum, and the introduced jet power does not exceed the Eddington luminosity of the SMBH, the blob radius will be strictly constrained by
\begin{equation}\label{RS}
\frac{R}{R_{\rm S}}\leqslant \frac{\sigma_{\rm pp}}{12\sigma_{\rm T}}\frac{\delta_{\rm D}^4}{\Gamma^4}\frac{L_{\rm Edd}}{L_{\rm TeV}^{\rm obs}},
\end{equation}
where $R_{\rm S}$ is the Schwarzschild radius of the SMBH and $L_{\rm TeV}^{\rm obs}$ is the EBL corrected TeV luminosity. It can be seen that  effective $pp$ interactions are more likely to occur in low TeV luminosity AGNs, otherwise the required $R$ will be too small. At present, LHAASO is the most sensitive equipment in VHE $\gamma$-ray band \citep{2019arXiv190502773C}. The LHAASO sensitivity curve consists of two components, the first is the water Cherenkov detector (WCDA), operating in the energies below 10 TeV, and the second is the KM2A array, sensitive to energies above 10 TeV \citep{2014APh....54...86C}. Since LHAASO is located in Sichuan Province, China, we select 12 low TeV luminosity jet-dominated AGNs in the northern sky as our sample. The detailed information of the sample is given in Table~\ref{table1}.

In this section, we apply the one-zone $pp$ model that proposed in this work to reproduce multi-wavelength SEDs of our sample. In the modeling, we do not aim to search for the best-fit model, but rather to show that $pp$ interactions in the framework of one-zone model can contribute to VHE emission that can be detected by LHAASO in the near future. Please note that steady-state EED and PED are used in the modeling, therefore low/quiescent state SEDs are studied in this work. In the following, we present brief descriptions and show fitting results for each source, respectively. The adopted free and derived/fixed parameters are shown in Table~\ref{parameters}.

\subsection{M 87}
As one of the closest AGNs, the Virgo Cluster galaxy M~87 harbors the first example of an extragalactic jet that has been observed by astronomers \citep{1918PLicO..13....9C}, hosts one of the most massive SMBH \citep{2019ApJ...875L...6E}, is detected as the first TeV RG \citep{2003A&A...403L...1A}, and is the first object to have a direct image of a SMBH ``shadow'' \citep{2019ApJ...875L...1E}. As a typical RG, the viewing angle of M~87 jet axis to the line of sight is between $15^\circ \sim 25^\circ$ \citep{1999ApJ...520..621B, 2009Sci...325..444A, 2018ApJ...855..128W}. The fastest variability timescale detected in VHE band is about 1 day, which constrains the size of the flaring emission region to a small scale \citep{2006Sci...314.1424A}. Since 2010, the VHE emission of M~87 is basically in a low-state \citep{2020MNRAS.492.5354M}. At present, the location of the VHE emitting region remains unclear. Various possible origins are discussed \citep{2007ApJ...663L..65C, 2007ApJ...671...85N, 2008MNRAS.385L..98T}, among which strong hints suggest that the variable VHE emission is from the core region \citep{2008ApJ...679..397A, 2009Sci...325..444A}.

\begin{figure}
\centering
\subfloat{
\includegraphics[width=0.94\columnwidth]{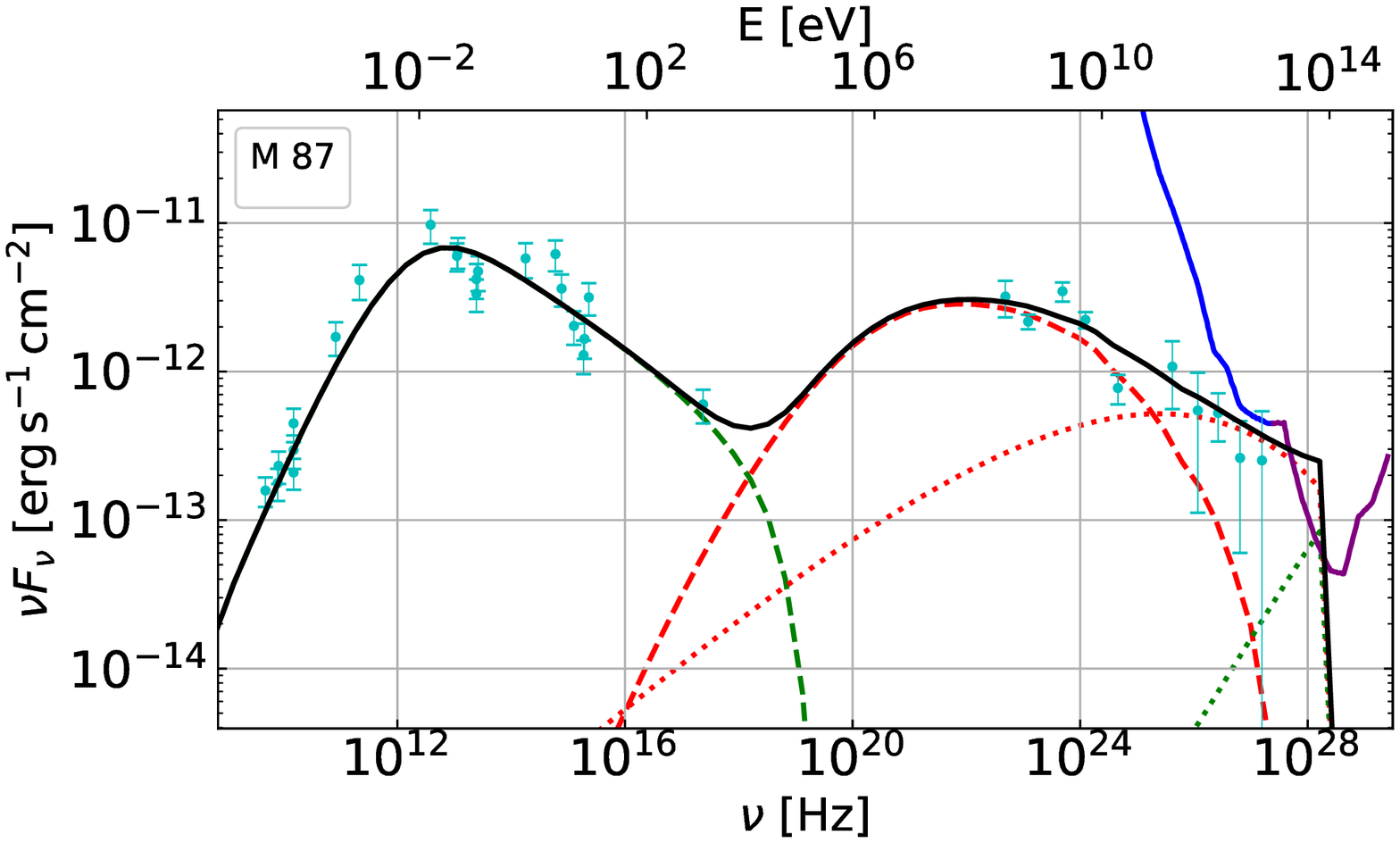}
}\hspace{-5mm}
\quad
\subfloat{
\includegraphics[width=0.94\columnwidth]{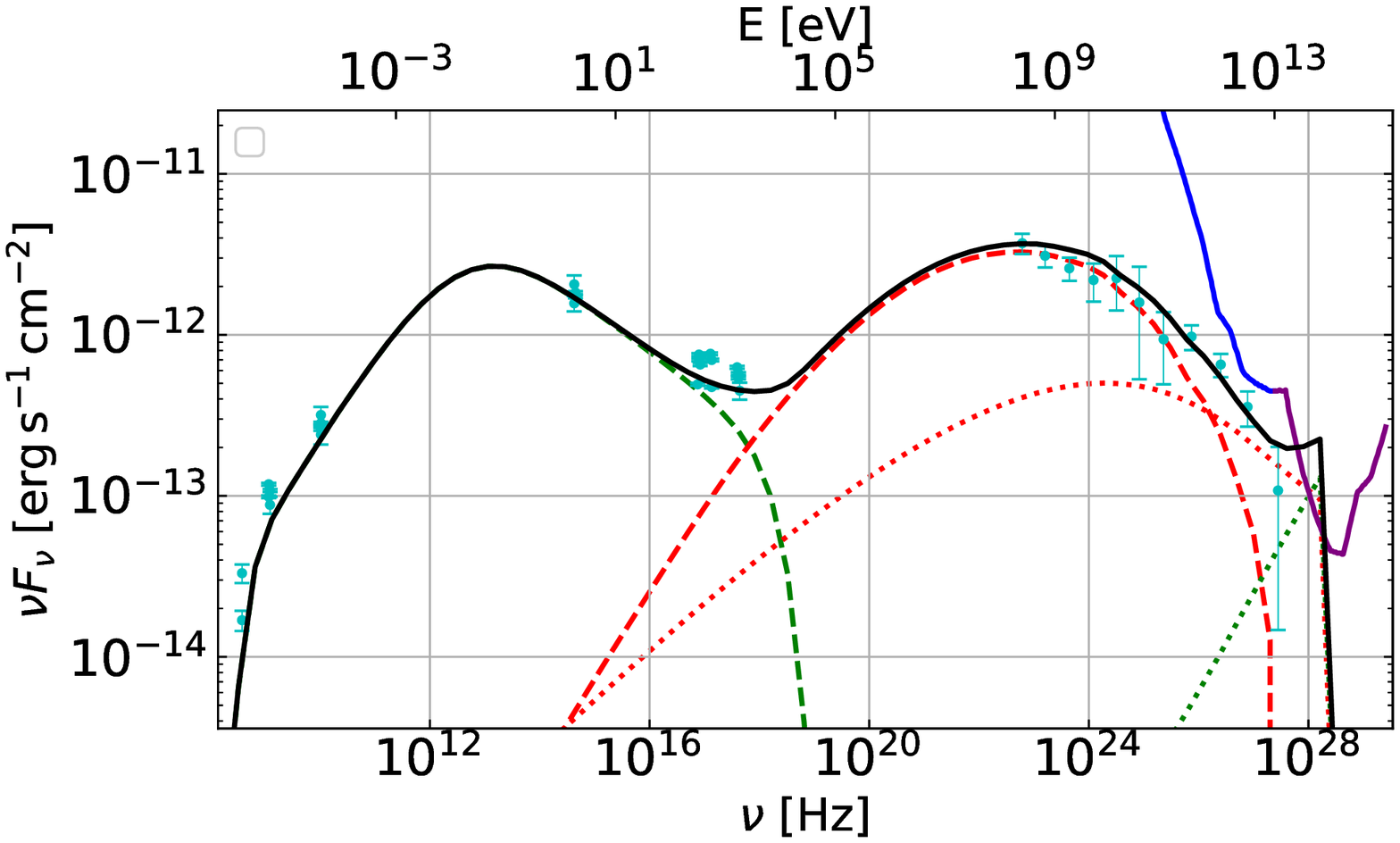}
}\hspace{-5mm}
\subfloat{
\includegraphics[width=0.94\columnwidth]{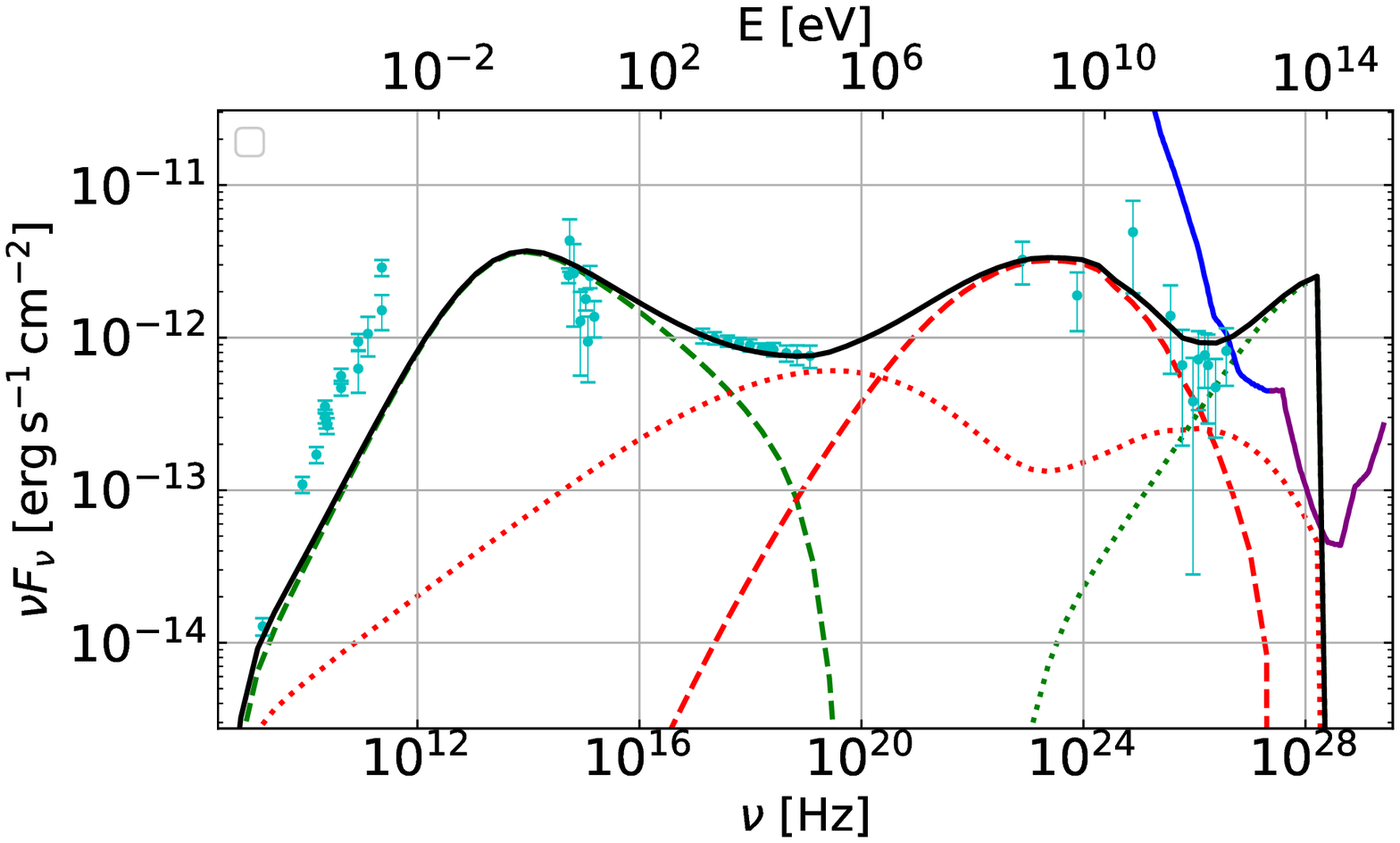}
}
\caption{The low-state multi-wavelength SEDs of M~87 core. The data points in the upper, middle, lower panels are taken from Ref.~\citep{2015MNRAS.450.4333D}, Ref.~\citep{2020MNRAS.492.5354M} and Ref.~\citep{2021ApJ...911L..11E}, respectively. The blue and purple solid curves represent the one-year sensitivities of WCDA and KM2A of LHAASO \citep{2019arXiv190502773C}, respectively . The dashed green and red curves represent the synchrotron and SSC emission from primary relativistic electrons in the blob, respectively. The dotted red curve represents the emission from pair cascades. The dotted green curve shows the $\gamma$-ray emission from $\pi_0$ decay in $pp$ interaction. The solid black curve is the total emission from the blob.
\label{M87}}
\end{figure}

For the core emission of M~87, Ref.~\citep{2015MNRAS.450.4333D} (upper panel) collects historical data to create a multi-wavelength SED for the average low-state, Ref.~\citep{2020MNRAS.492.5354M} (middle panel) provides a multi-wavelength SED between 2012 and 2015 during the low VHE $\gamma$-ray state, and Ref.~\citep{2021ApJ...911L..11E} (lower panel) obtains a low-state multi-wavelength SED during the 2017 Event Horizon Telescope (EHT) campaign. As shown in Figure~\ref{M87}, these three low-state SEDs are reproduced with the one-zone $pp$ model. In the modeling, we fix $\theta_{\rm obs} = 15^\circ$ and $\Gamma=3$, which are consistent with apparent motion observations \citep{1999ApJ...520..621B, 2007ApJ...660..200L, 2013ApJ...774L..21M}. It can be seen that the broadband spectrum from radio to GeV band is contributed by the leptonic emission from the primary electrons (dashed curves) and VHE spectrum is dominated by the emission from $pp$ interactions (dotted curves). Despite the production site of VHE emission remains unclear, our modeling results suggest that $pp$ interactions occur in the core region can contribute to the low-state VHE spectrum. Under the current parameter set, the model predicted $\gamma$-ray flux well exceeds the one-year sensitivity of KM2A array of LHAASO. Therefore, $\gamma$-ray emission from M~87 might be detected by LHAASO in the near future, which could test if efficient $pp$ interactions do exist in the jet of M~87. Please note that when fitting the SED during the 2017 EHT campaign \citep{2021ApJ...911L..11E}, the model predicted $\gamma$-ray flux is very high, which also exceeds the one-year sensitivity of WCDA of LHAASO, since its relative flat VHE spectrum has a looser constraint compared to those of Ref.~\citep{2015MNRAS.450.4333D} and Ref.~\citep{2020MNRAS.492.5354M}. These three low-state SEDs are similar to each other, however the SED of Ref.~\citep{2021ApJ...911L..11E} exhibits a significant `flat' X-ray excess, which is hard to be explained by leptonic emission alone (as shown in Figures 17 and 18 of Ref.~\citep{2021ApJ...911L..11E}). In our modeling, we suggest that it can be well fitted by the cascades emission (also see Ref.~\citep{2022arXiv220814756B} for the disc origin explanation). In order to do so, compared to free parameters adopted in the fitting of upper and middle panels, a relative lower $\gamma_{\rm p,max}$ is required. At the same time, if secondary cascades with lower-energies can generate enough emission to explain the X-ray excess, a stronger magnetic field also needs to be introduced. Moreover, in order to make the SSC emission of primary electrons still fit GeV data contemporaneously, a relatively smaller blob radius has to be set. From Equation \ref{n_H}, it can be seen that the number density of cold protons will be increased accordingly. In general, we believe that the `flat' X-ray excess given by Ref.~\citep{2021ApJ...911L..11E} can be well explained by cascades emission, and implies that the emitting region will be closer to the jet base. On the other hand, we do not fit the radio spectrum when reproducing the SED of Ref.~\citep{2021ApJ...911L..11E}, because this well observed radio spectrum exhibits an obvious core-shift feature, i.e., higher frequency radio emission comes from a denser region, which is more appropriate to be explained by a conical jet model \citep{1979ApJ...232...34B} rather than the simplified one-zone model. Radio observation \citep{2021ApJ...911L..11E} suggests that the lowest frequency data point comes from a region with radius less than $650~R_{\rm S}$, i.e., $\leqslant 1.2\times10^{18}\rm~cm$, which is consistent with the blob radius set during the fitting. 

\subsection{IC 310}
The radio galaxy IC 310, also known as B0313+411 and J0316+4119, located on the outskirts of Perseus Cluster. With a viewing angle of $10^\circ-20^\circ$ \citep{2014Sci...346.1080A}, IC 310 is considered as a transition object at the borderline dividing low-luminosity FR I RGs and BL Lacs \citep{2012A&A...538L...1K, 2014A&A...563A..91A}. Its VHE emission with an extreme short timescale ($\sim 5\rm min$) is firstly detected by MAGIC \citep{2014Sci...346.1080A}.  

\begin{figure}
\centering
\subfloat{
\includegraphics[width=0.95\columnwidth]{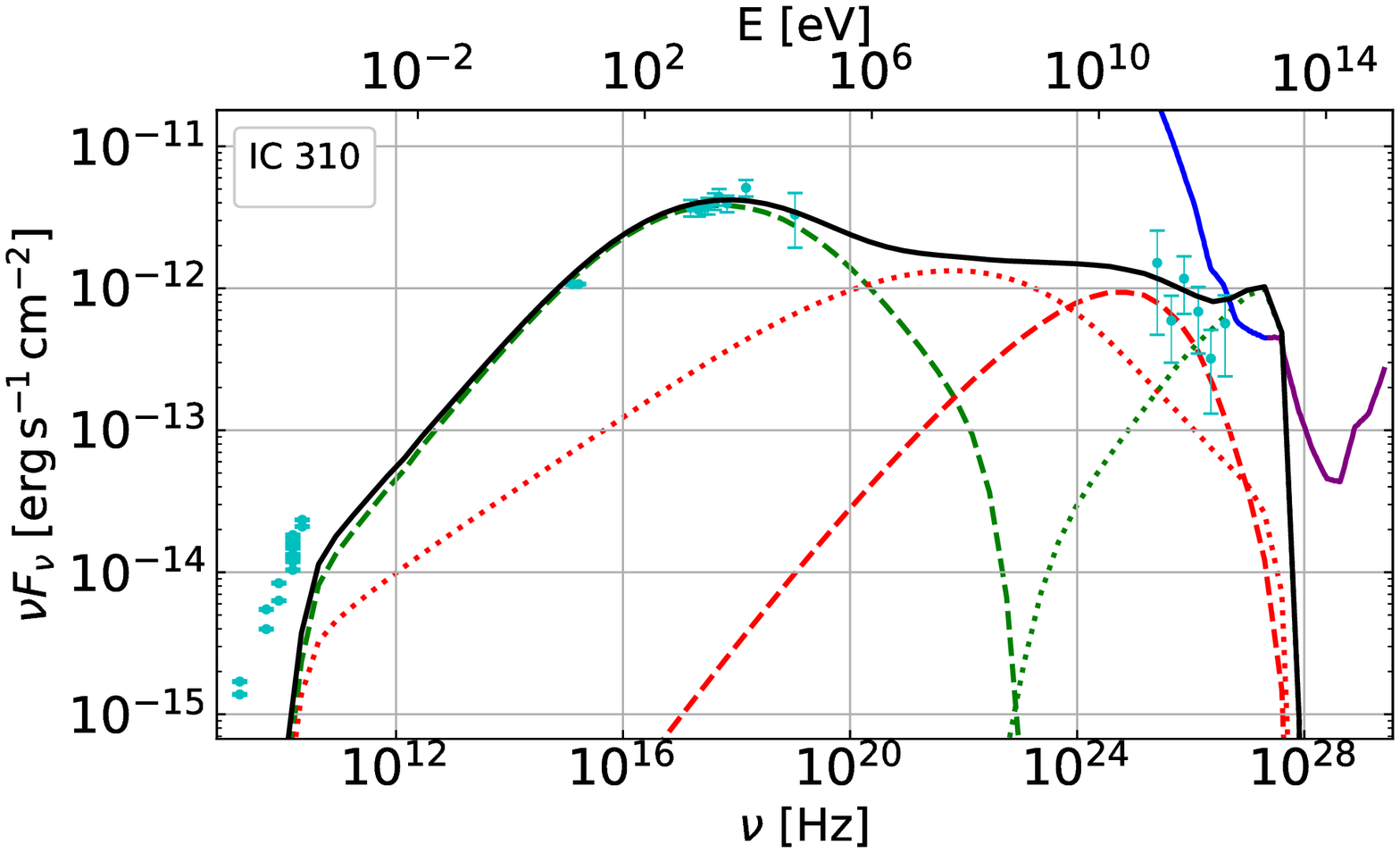}
}\hspace{-5mm}
\quad
\subfloat{
\includegraphics[width=0.95\columnwidth]{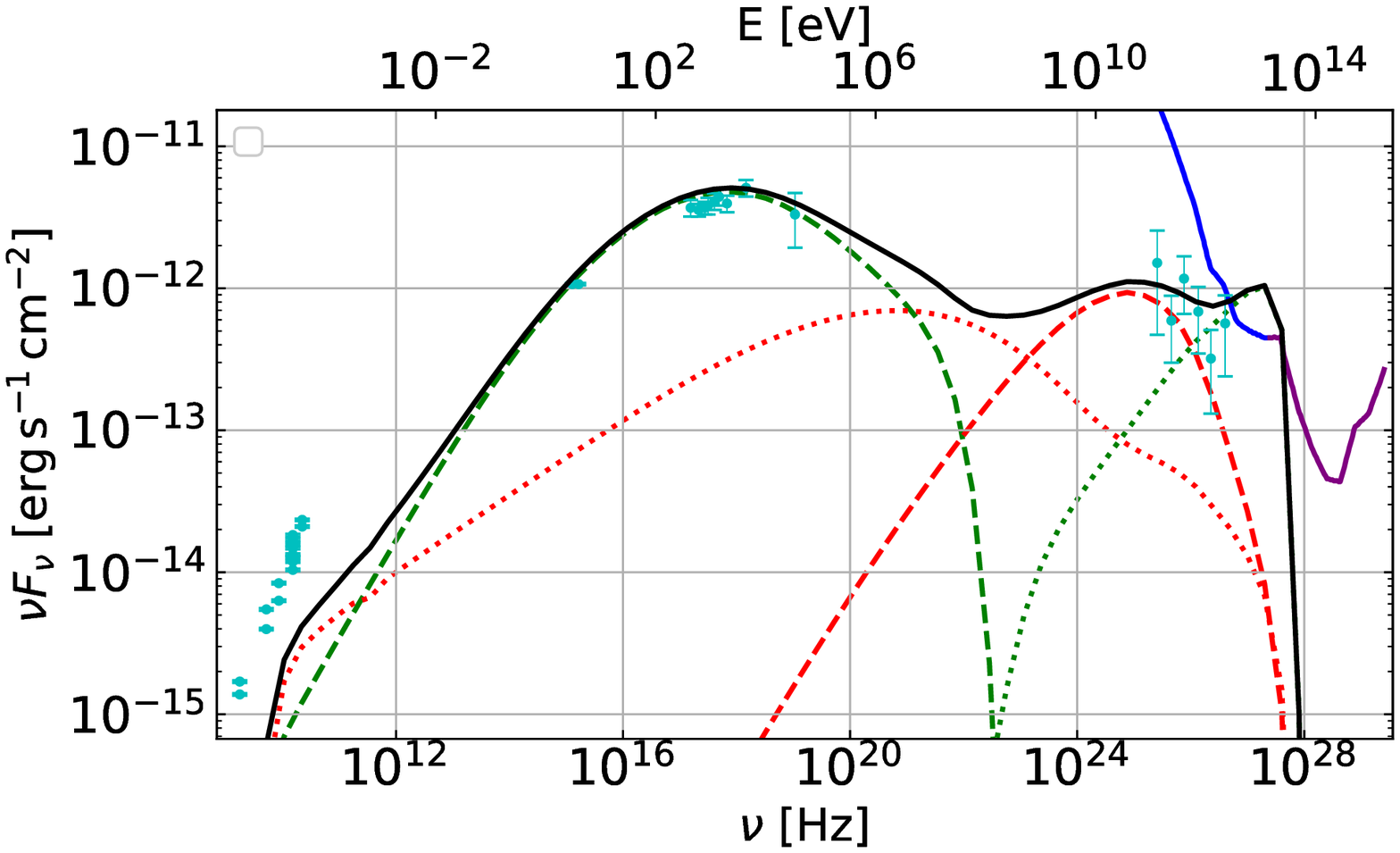}
}
\caption{The low-state multi-wavelength SEDs of IC 310 that taken from Ref.~\citep{2017AA...603A..25A}. The upper panel shows the fitting result when assuming IC 310 is a blazar with $\theta_{\rm obs}=10^{\circ}$. The lower panel shows the fitting result when assuming IC 310 is a FR I RG with $\theta_{\rm obs}=20^{\circ}$. The line styles have the same meaning as in Figure~\ref{M87}.
\label{IC310}}
\end{figure}

Ref.~\citep{2017AA...603A..25A} provides the first simultaneous SED during the multi-wavelength campaign from November 2012  to January 2013 when the VHE emission is in a low-state. Since IC 310 is a transition AGN, Ref.~\citep{2017AA...603A..25A} reproduce the SED in two cases, which assume that IC 310 is a blazar (upper panel) or a RG (lower panel). Following Ref.~\citep{2017AA...603A..25A}, we also fit the SED by treating it as a blazar with $\theta_{\rm obs}=10^\circ$ and a RG with $\theta_{\rm obs}=20^\circ$, respectively. Our fitting results are given in Figure~\ref{IC310}. For the leptonic emission, our fitting results and adopted parameters are basically consistent with those in Ref.~\citep{2017AA...603A..25A}, although there are some slight differences. In our modeling, the VHE spectrum is explained by the $\pi^0$ decay, which also have a considerable contribution on the energy band that exceeds the one-year sensitivity of LHAASO.

\begin{table*}
\caption{Parameters for SED Fitting with the One-zone $pp$ Model}
\centering
\resizebox{\textwidth}{62mm}{
\begin{tabular}{cccccccccccc}
\hline\hline
Free parameters	&	panel	&	$\Gamma$	&	$R$~(cm)	&	$B$~(G)	&	$L_{\rm e,inj}~(\rm erg~s^{-1})$	&	$\gamma_{\rm e,b}$	&	$\alpha_{\rm e,1}$	&	$\alpha_{\rm e,2}$	&	$\xi$	&	$\gamma_{\rm p,max}$	&	$\alpha_{\rm p}$	\\
Derived/Fixed parameters	&		&	$\theta_{\rm obs}~(\circ)$	&	$\delta_{\rm D}$	&	$\gamma_{\rm e,min}$	&	$\gamma_{\rm e,max}$	&	$\gamma_{\rm p,min}$	&	$\chi_{\rm p}$	&	$L_{\rm Edd}~(\rm erg~s^{-1})$	&	$n_{\rm H}~(\rm cm^{-3})$	&	$f_{\rm pp}$	&	$\chi^2$	\\
\hline																							
M 87 	&	upper panel	&	3	&	$6\times10^{17}$	&	$2.5\times10^{-3}$	&	$7\times10^{42}$	&	$8\times10^3$	&	1.2	&	3.5	&	$5\times10^{-3}$	&	$4.1\times10^7$	&	1.5	\\
	&		&	15	&	3.7	&	1	&	$10^7$	&	1	&	0.5	&	$9.0\times10^{47}$	&	$2.0\times10^{3}$	&	$1.8\times10^{-5}$	&	2.9	\\
	&	middle panel	&	3	&	$6\times10^{17}$	&	$1\times10^{-3}$	&	$3\times10^{43}$	&	$2\times10^4$	&	1.9	&	3.5	&	$5\times10^{-3}$	&	$1.7\times10^7$	&	1.5	\\
	&		&	15	&	3.7	&	1	&	$10^7$	&	1	&	0.5	&	$9.0\times10^{47}$	&	$2.0\times10^{3}$	&	$1.8\times10^{-5}$	&	9.3	\\
	&	lower panel	&	3	&	$1.5\times10^{17}$	&	$5\times10^{-3}$	&	$2.2\times10^{42}$	&	$2\times10^4$	&	1.5	&	3.5	&	$2.4\times10^{-3}$	&	$2.1\times10^6$	&	1.5	\\
	&		&	15	&	3.7	&	1	&	$10^7$	&	1	&	0.5	&	$9.0\times10^{47}$	&	$3.1\times10^{4}$	&	$7.1\times10^{-5}$	&	1.9	\\
IC 310 	&	upper panel	&	5	&	$4\times10^{15}$	&	0.3	&	$1.3\times10^{41}$	&	$1.5\times10^6$	&	2	&	3.1	&	0.14	&	$3.3\times10^6$	&	1.5	\\
	&		&	10	&	5.7	&	1	&	$10^7$	&	1	&	0.5	&	$1.3\times10^{47}$	&	$2.3\times10^{6}$	&	$1.4\times10^{-4}$	&	5.1	\\
	&	lower panel	&	3	&	$2\times10^{16}$	&	0.17	&	$7\times10^{41}$	&	$1.1\times10^6$	&	1.6	&	2.8	&	1	&	$1.9\times10^6$	&	1.5	\\
	&		&	20	&	2.9	&	1	&	$10^7$	&	1	&	0.5	&	$1.3\times10^{47}$	&	$1.4\times10^{5}$	&	$4.1\times10^{-5}$	&	4.3	\\
3C 264	&		&	10	&	$4\times10^{15}$	&	0.1	&	$5.3\times10^{40}$	&	$6.9\times10^3$	&	1.4	&	2.9	&	1	&	$9.9\times10^6$	&	1.5	\\
	&		&	5	&	11.4	&	1	&	$10^7$	&	1	&	0.5	&	$6.9\times10^{46}$	&	$1.5\times10^{5}$	&	$1.1\times10^{-5}$	&	3.1	\\
Mrk 421 	&		&	40	&	$5\times10^{15}$	&	0.1	&	$4\times10^{40}$	&	$3.1\times10^5$	&	2.4	&	6.5	&	1	&	$1.4\times10^{5}$	&	1.5	\\
	&		&		&	40	&	1	&	$10^7$	&	1	&	0.5	&	$1.9\times10^{47}$	&	$1.6\times10^{4}$	&	$1.23\times10^{-6}$	&	6.1	\\
Mrk 501 	&		&	35	&	$8\times10^{14}$	&	0.3	&	$6.2\times10^{40}$	&	$1.1\times10^5$	&	2.2	&	3.5	&	1	&	$5.8\times10^{5}$	&	1.5	\\
	&		&		&	35	&	1	&	$10^7$	&	1	&	0.5	&	$5.8\times10^{46}$	&	$3.4\times10^{5}$	&	$5.4\times10^{-6}$	&	2.2	\\
1ES 2344+514 	&		&	30	&	$1.2\times10^{15}$	&	0.2	&	$2.9\times10^{40}$	&	$5\times10^4$	&	2	&	4	&	1	&	$6.1\times10^{5}$	&	1.5	\\
	&		&		&	30	&	1	&	$10^7$	&	1	&	0.5	&	$8.7\times10^{46}$	&	$1.4\times10^{6}$	&	$2.1\times10^{-5}$	&	2.6	\\
TXS 0210+515 	&		&	10	&	$1.1\times10^{15}$	&	0.65	&	$3.1\times10^{40}$	&	$1.5\times10^5$	&	1.5	&	2	&	0.3	&	$1.8\times10^{6}$	&	1.5	\\
	&		&		&	10	&	1	&	$10^7$	&	1	&	0.5	&	$1.3\times10^{47}$	&	$3.6\times10^{6}$	&	$5.4\times10^{-5}$	&	1.9	\\
1ES 2037+521 	&		&	30	&	$1.1\times10^{15}$	&	0.07	&	$7.1\times10^{39}$	&	$5.8\times10^5$	&	1.5	&	4	&	1	&	$1.9\times10^{5}$	&	1.5	\\
	&		&		&	30	&	1	&	$10^7$	&	1	&	0.5	&	$1.3\times10^{47}$	&	$2.2\times10^{6}$	&	$3.2\times10^{-5}$	&	1.1	\\
RGB J0152+017 	&		&	30	&	$1.1\times10^{15}$	&	0.1	&	$1.5\times10^{40}$	&	$1.3\times10^5$	&	1.5	&	3.7	&	1	&	$2.8\times10^{4}$	&	1.5	\\
	&		&		&	30	&	1	&	$10^7$	&	1	&	0.5	&	$1.3\times10^{47}$	&	$5.4\times10^{5}$	&	$8.12\times10^{-6}$	&	6.7	\\
1ES 1741+196 	&		&	30	&	$1.3\times10^{15}$	&	0.19	&	$5\times10^{40}$	&	$4\times10^5$	&	2.2	&	4	&	1	&	$5.3\times10^{5}$	&	1.5	\\
	&		&		&	30	&	1	&	$10^7$	&	1	&	0.5	&	$1.3\times10^{47}$	&	$2.2\times10^{6}$	&	$3.2\times10^{-5}$	&	2.9	\\
RGB J2042+244 	&		&	30	&	$7\times10^{14}$	&	0.6	&	$8\times10^{39}$	&	$4\times10^4$	&	1.5	&	3.5	&	1	&	$1.2\times10^{6}$	&	1.7	\\
	&		&		&	30	&	1	&	$10^7$	&	1	&	0.5	&	$1.3\times10^{47}$	&	$4.4\times10^{6}$	&	$4.7\times10^{-5}$	&	2.5	\\
1ES 0229+200 	&		&	30	&	$2.7\times10^{14}$	&	1.3	&	$6.8\times10^{39}$	&	$2.3\times10^5$	&	1.3	&	4	&	1	&	$9.7\times10^{4}$	&	1.5	\\
	&		&		&	30	&	1	&	$10^7$	&	1	&	0.5	&	$2.5\times10^{47}$	&	$5.5\times10^{7}$	&	$2.2\times10^{-4}$	&	2.6	\\
\hline
\label{parameters}
\end{tabular}}
\textbf{Notes.} The blob radius $R$ is not a completely free parameter. In order to make the emission from $pp$ interactions contribute to VHE and UHE bands, the adopted values of $R$ are lower than the maximum value constrained by Eq.~(\ref{RS}). As suggested in Ref.~\citep{2022A&A...659A.184L}, we fix $\chi_{\rm p}=0.5$. The maximum parameter space can be obtained under this condition, which also indicates that the kinetic power of cold protons and the relativistic proton injection power each account for half of the total jet power. The number density of cold protons $n_{\rm H}$ is derived by Eq.~(\ref{n_H}) and the $pp$ interaction efficiency $f_{\rm pp}$ is obtained by setting a constant cross section for inelastic $pp$ interactions $\sigma_{\rm pp}=6\times10^{-26}~\rm cm^2$. The corresponding chi-square $\chi^2$ value for each object is calculated by $\chi^2 = \frac{1}{m-dof}\sum_{i=1}^{m}(\frac{\hat {y}_i-y_i}{\sigma_i})^2$, where $m$ is the number of quasi-simultaneous observational data points, $dof$ are the degrees of freedom, $\hat {y}_i$ are the expected values from the model, $y_i$ are the observed data and $\sigma_i$ is the standard deviation for each data point.
\end{table*}

\subsection{3C 264}
\begin{figure}
\centering
\includegraphics[width=0.95\columnwidth]{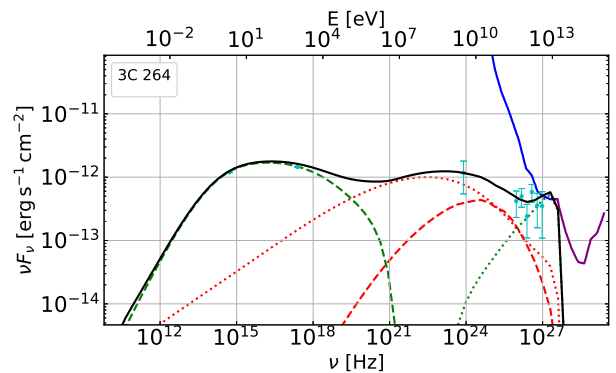}
\caption{The low-state multi-wavelength SED of 3C 264 that taken from Ref.~\citep{2020ApJ...896...41A}. The line styles have the same meaning as in Figure~\ref{M87}.
\label{3C264}}
\end{figure}

The FR I RG 3C 264 is the fourth RG detected at VHE band \citep{2020ApJ...896...41A}. In 2017-2019, a mild VHE variability is detected by VERITAS, while other energy bands are basically in the low-state. Although 3C 264 is classified as a RG, the observation of apparent speeds constrains its jet viewing angle to a small value $\theta_{\rm obs}<10^{\circ}$ \citep{2019A&A...627A..89B}. Treating 3C 264 as a BL Lac, Ref.~\citep{2020ApJ...896...41A} reproduce its simultaneous SED by setting $\delta_{\rm D}=10$. Our fitting result is shown in Figure~\ref{3C264}. In our modeling, we also treat 3C 264 as a blazar, and fix $\theta_{\rm obs} = 5$ and $\Gamma=10$. Compared to the one-zone leptonic modeling in Ref.~\citep{2020ApJ...896...41A}, our modeling improves the fitting of VHE spectrum by further considering the emission from $pp$ interactions. Since 3C 264 is the most distant RG, the EBL absorption becomes significant. Constrained by the TeV data points, the model predicted $\gamma$-ray flux generated in the $\pi^0$ decay slightly exceeds the one-year sensitivity of WCDA of LHAASO.

\subsection{Mrk 421}
As one of best studied blazars, Mrk 421 is also the first VHE emitter detected by Whipple telescope in 1992 \citep{1992Natur.358..477P}. Its multi-wavelength light curves are highly variable and show complicated variability patterns. For instance, X-ray variability is normally correlate to the VHE variability \citep{2008ApJ...677..906F, 2011ApJ...738...25A, 2013PASJ...65..109C, 2015A&A...578A..22A}, however the correlations of the variabilities between other bands and these two bands are reported weak or inexistent \citep{1995ApJ...449L..99M, 2013PASJ...65..109C, 2007ApJ...663..125A, 2016ApJ...819..156B}.

\begin{figure}
\centering
\includegraphics[width=0.95\columnwidth]{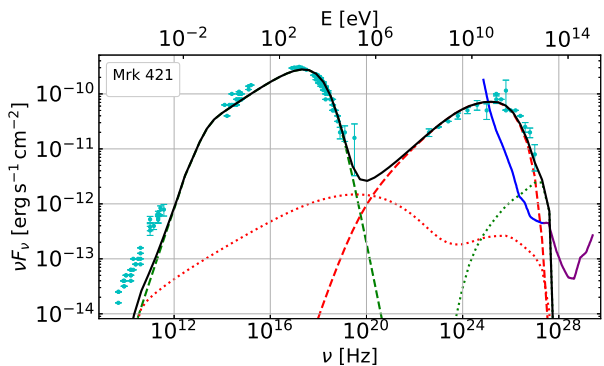}
\caption{The low-state multi-wavelength SED of Mrk 421 that taken from Ref.~\citep{2011ApJ...736..131A}. The line styles have the same meaning as in Figure~\ref{M87}.
\label{421}}
\end{figure}

The averaged SED of Mrk 421 resulting from quasi-simultaneous observations integrated over a period of 4.5 months is provided by Ref.~\citep{2011ApJ...736..131A}, which is the most complete SED ever collected. In our fitting result, as shown in Figure~\ref{421}, the multi-wavelength SED is interpreted by the leptonic emission from primary electrons. The hadronic emission from $\pi^0$ decay has a sub-dominant contribution to the highest energy TeV data point, and has a significant contribution on the higher energy band, which exceeds the one-year sensitivity of LHAASO. It should be noted that the flux of the current detected VHE data points far exceeds the one-year sensitivity of WCDA of LHAASO, therefore regardless of the origin of the currently observed VHE spectrum, higher energy emission is likely to be discovered by LHAASO because the current VHE spectrum does not exhibit truncated feature. From our fitting result, it can be seen that the sub-dominant hadronic emission harden the TeV spectrum to a certain degree compared to the SSC emission under the KN regime. If LHAASO can give constraints on the spectral shape of higher-energy VHE spectrum in the future, it may help determine whether there are additional emission components, such as the hadronic emission shown in this work.  In addition, cascade emission also has a sub-dominant contribution to the hard X-ray excess \citep{2016ApJ...827...55K, 2017ApJ...842..129C}. If the injection luminosity of relativistic protons or the number density of cold protons increase for a short period of time, the correlated flares of hard X-ray and VHE bands may arise because they both originate from $pp$ interactions. While correlated flares may not occur in other bands since they are from the leptonic processes. 

\subsection{Mrk 501}
\begin{figure}
\centering
\includegraphics[width=0.95\columnwidth]{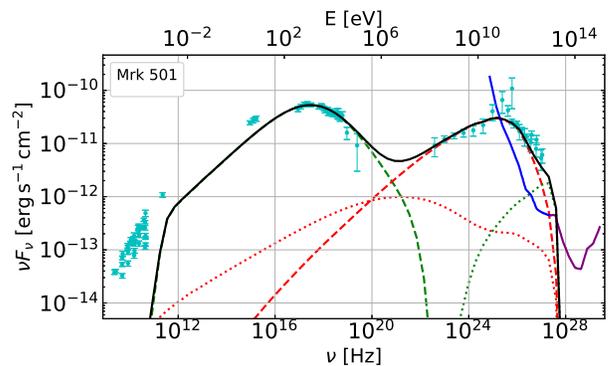}
\caption{The low-state multi-wavelength SED of Mrk 501 that taken from Ref.~\citep{2011ApJ...727..129A}. The line styles have the same meaning as in Figure~\ref{M87}.
\label{501}}
\end{figure}

After Mrk 421, Mrk 501 is the second extragalactic source discovered at VHE band in 1995 \citep{1996ApJ...456L..83Q, 1997A&A...320L...5B}, and it has been intensively studied in the past. Similar to Mrk 421, a detailed low-state multi-wavelength SED obtained during 4.5 months campaign is given by Ref.~\citep{2011ApJ...727..129A}. Our fitting result is shown in Figure~\ref{501}. Similar to the modeling of Mrk 421, the multi-wavelength SED of Mrk 501 is also interpreted by the leptonic emission from primary electrons, while the hadronic emission from $pp$ interactions has a sub-dominant contribution to the highest energy TeV data point, and has a significant contribution on the higher energy band, which exceeds the one-year sensitivity of LHAASO. It is worth noting that the flux of its observed low-state VHE spectrum also far exceeds the one-year sensitivity of WCDA of LHAASO.  Therefore, similar to the VHE emission of Mrk 421, LHAASO should be able to detect higher energy TeV emission and give constraints on the spectral shape in the future.

\subsection{1ES 2344+514}
\begin{figure}
\centering
\includegraphics[width=0.95\columnwidth]{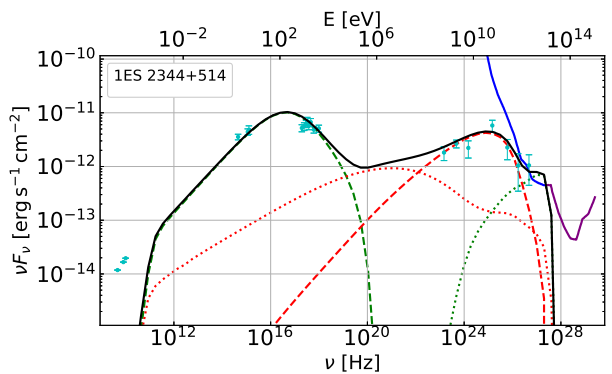}
\caption{The low-state multi-wavelength SED of 1ES 2344+514 that taken from Ref.~\citep{2013AA...556A..67A}. The line styles have the same meaning as in Figure~\ref{M87}.
\label{2344}}
\end{figure}

1ES 2344+514 is the third extragalactic source discovered at VHE band after Mrk 421 and Mrk 501 \citep{1998ApJ...501..616C}. The day-scale flares in VHE band are usually detected \citep{1998ApJ...501..616C, 2011ApJ...738..169A}. The first low-state simultaneous radio to VHE observations of 1ES 2344+514 is presented by Ref.~\citep{2013AA...556A..67A}. In Figure~\ref{2344}, we reproduce its SED with the one-zone $pp$ model. In our modeling, the multi-wavelength emission is dominated by the leptonic emission from primary electrons. The emission from $pp$ interactions has a significant contribution to the highest energy data point, which improves the fitting of VHE spectrum and exceed the one-year sensitivity of WCDA of LHAASO. Due to the strong EBL absorption, the higher energy TeV spectrum from $\pi^0$ decay is truncated so that it cannot be detected by KM2A.

\subsection{RGB J0152+017}
\begin{figure}
\centering
\includegraphics[width=0.95\columnwidth]{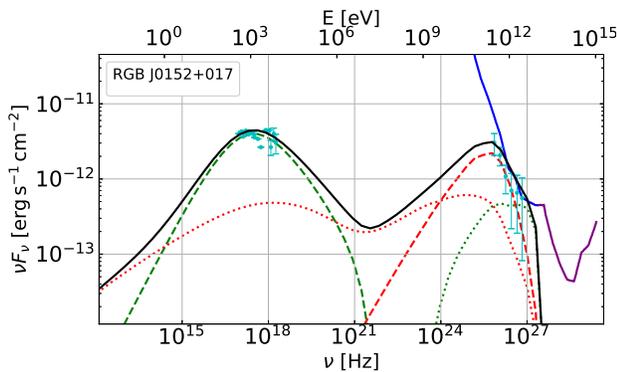}
\caption{The low-state multi-wavelength SED of RGB J0152+017 that taken from Ref.~\citep{2008AA...481L.103A}. The line styles have the same meaning as in Figure~\ref{M87}.
\label{RGB0152}}
\end{figure}

The VHE emission of RGB J0152+017 is discovered by HESS at the end of 2007 \citep{2008AA...481L.103A}. Ref.~\citep{2008AA...481L.103A} provides the only simultaneous multi-wavelength SED that can be found in literature, and multi-wavelength light curves do not show any significant variability. Our fitting result is given in Figure~\ref{RGB0152}. The SED is well fitted by leptonic emission from primary electrons, and emission from $\pi^0$ decay has a sub-dominant contribution at TeV band, improving the fitting of the TeV spectrum to a certain extent. Because of the significant EBL absorption, the flux of model predicted TeV emission decreases rapidly, nevertheless still slightly exceeds the one-year sensitivity of WCDA of LHAASO.

\subsection{1ES 1741+196}
\begin{figure}
\centering
\includegraphics[width=0.95\columnwidth]{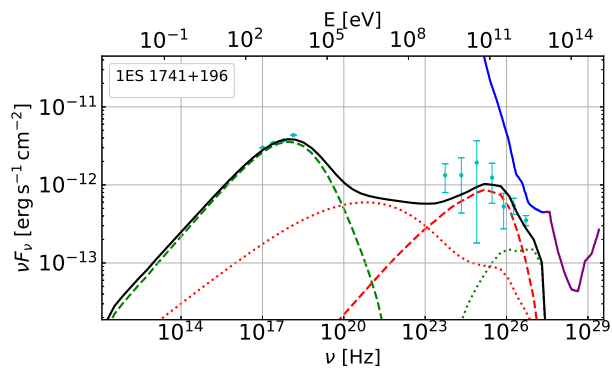}
\caption{The low-state multi-wavelength SED of 1ES 1741+196 that taken from Ref.~\citep{2017MNRAS.468.1534A}. The line styles have the same meaning as in Figure~\ref{M87}.
\label{1741}}
\end{figure}

1ES 1741+196 is discovered as a VHE emitter by MAGIC. The observation results of MAGIC and other wavebands are provided by Ref.~\citep{2017MNRAS.468.1534A}. No significant variabilities are found in the multi-wavelength light curves. In Figure~\ref{1741}, we reproduce its SED with the one-zone $pp$ model. Similar to the fitting result of RGB J0152+017, the multi-wavelength emission is dominated by the leptonic processes, and hadronic emission only has a contribution on the highest energy TeV data point, but does not exceed the one-year sensitivity of LHAASO because of the severe EBL absorption.

\subsection{1ES 0229+200}
\begin{figure}
\centering
\includegraphics[width=0.95\columnwidth]{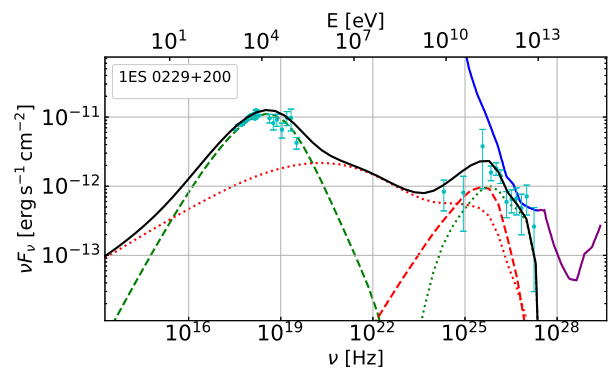}
\caption{The low-state multi-wavelength SED of 1ES 0229+200 that taken from Ref.~\citep{2014ApJ...782...13A}. The line styles have the same meaning as in Figure~\ref{M87}.
\label{0229}}
\end{figure}

1ES 0229+200 is the prototype of extreme HSP blazars ($\nu_{\rm S,peak}\gtrsim10^{17}~\rm Hz$; \citep{2001A&A...371..512C, 2018MNRAS.480.2165A}). The most remarkable feature of 1ES 0229+200 is the hard TeV spectrum, which poses a challenge to the conventional one-zone model \citep{2009MNRAS.399L..59T, 2011A&A...534A.130K, 2019ApJ...871...81X}. Analytical calculations suggest that $pp$ interactions may have the potential to explain the hard TeV spectrum of 1ES 0229+200 \citep{2022A&A...659A.184L}. In this work, we include 1ES 0229+200 in our sample, and use numerical model to reproduce its multi-wavelength SED, especially the hard-TeV spectrum \citep{2014ApJ...782...13A}. The fitting result is given in Figure~\ref{0229}. It can be seen that the hard-TeV spectrum is well explained by the $\gamma$-ray generated from $\pi^0$ decay, although a very compact emitting region is introduced, just as indicated by Ref.~\citep{2022A&A...659A.184L}. At present, no evidence of fast variability in the VHE band of 1ES 0229+200 is found, therefore the injection of relativistic protons in such a compact blob have to be continuous. Due to the significant EBL absorption, the TeV emission generated in $\pi^0$ decay does not exceed the one-year sensitivity of LHAASO. On the other hand, since the currently detected TeV data points are close to the LHAASO sensitivity, LHAASO may be able to detecte its high-state TeV emission in the future.

\subsection{TXS 0210+515, 1ES 2037+521, RGB J2042+244}
\begin{figure}
\centering
\subfloat{
\includegraphics[width=0.95\columnwidth]{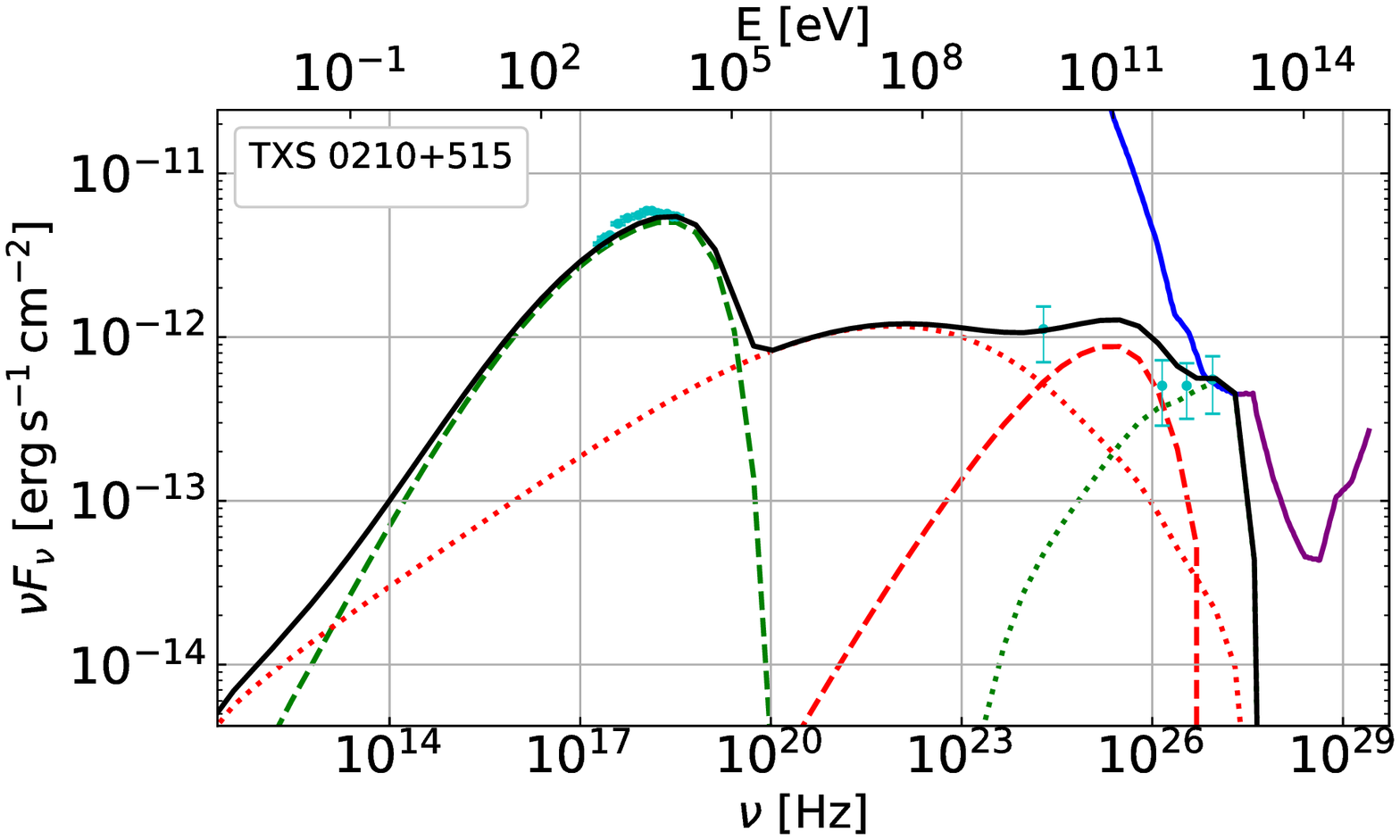}
}\hspace{-5mm}
\quad
\subfloat{
\includegraphics[width=0.95\columnwidth]{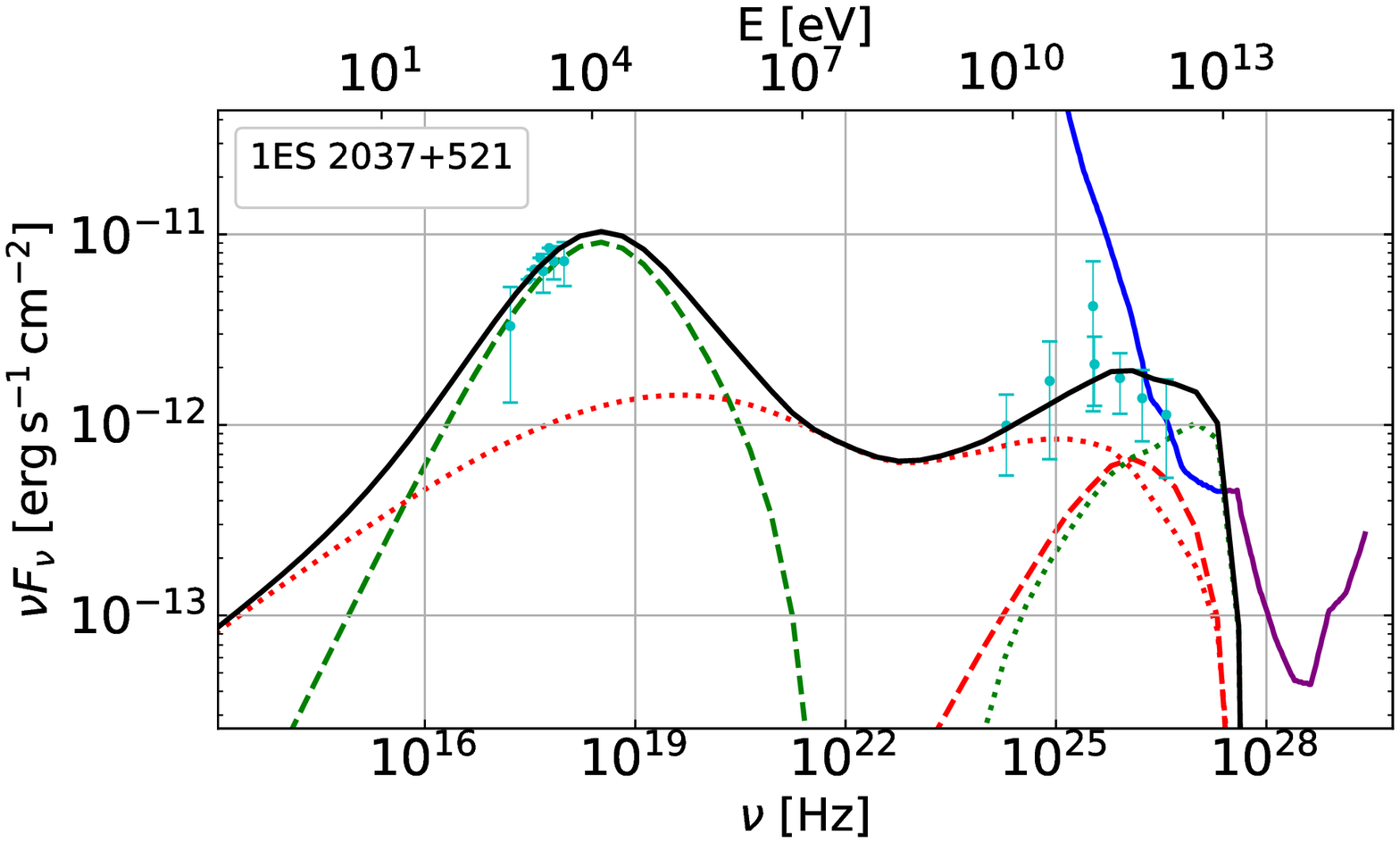}
}\hspace{-5mm}
\subfloat{
\includegraphics[width=0.95\columnwidth]{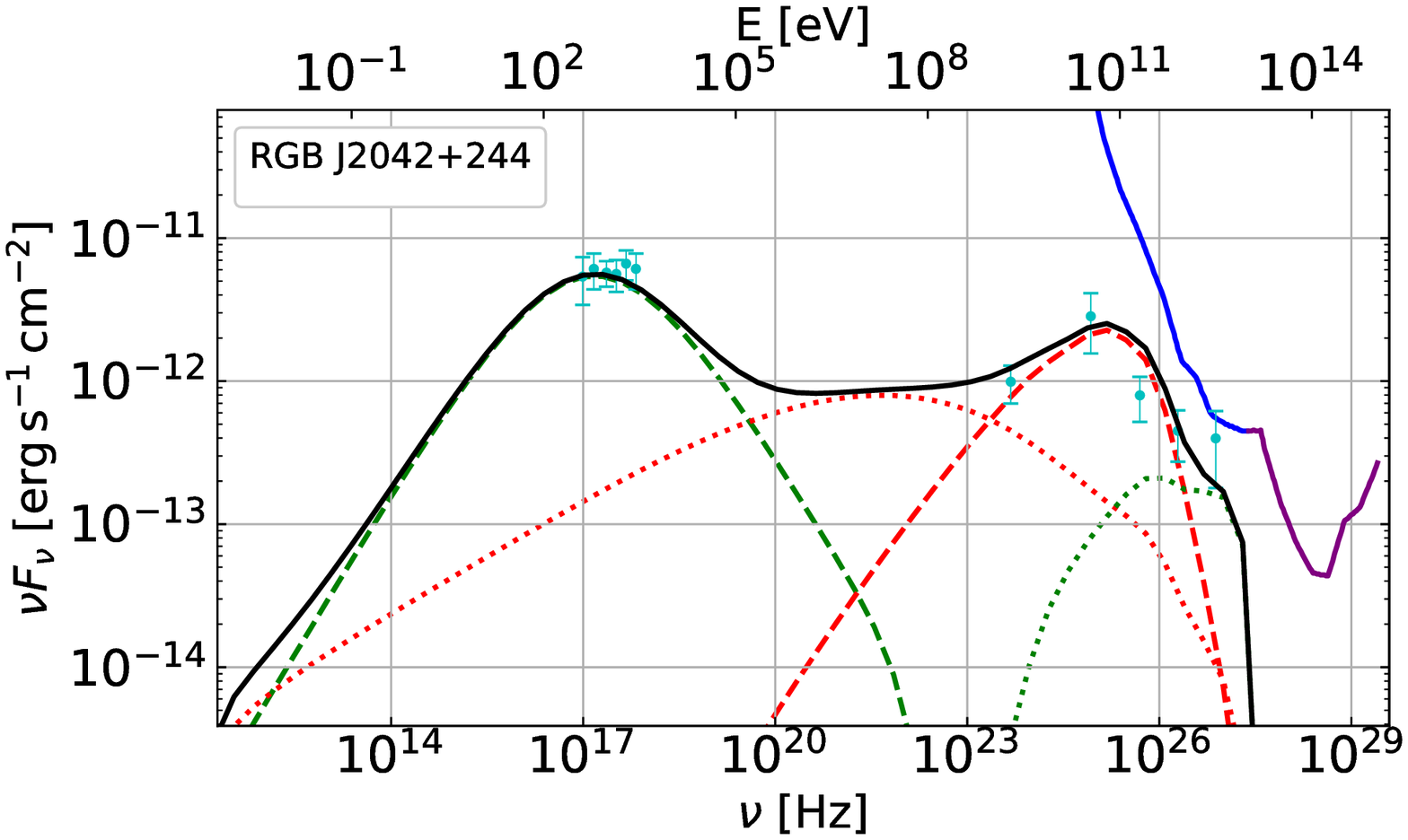}
}
\caption{The low-state multi-wavelength SEDs of TXS 0210+515, 1ES 2037+521 and RGB J2042+244 that taken from Ref.~\citep{2020ApJS..247...16A}. The line styles have the same meaning as in Figure~\ref{M87}.
\label{hard}}
\end{figure}

Recently, Ref.~\citep{2020ApJS..247...16A} reports a new sample of hard-TeV blazars, and collects their simultaneous multi-wavelength SEDs. Among them, three sources, i.e., TXS 0210+515, 1ES 2037+521 and RGB J2042+244, with relative low TeV luminosities are selected in our sample. Our fitting results are shown in Figure~\ref{hard}. In the modeling, the major contributors to multi-wavelength emission are leptonic processes. Compared to the fitting results of the conical jet model, spine-layer model and proton synchrotron model that given in Ref.~\citep{2020ApJS..247...16A}, the $\gamma$-ray spectra generated in the $\pi^0$ decay improve the fitting of hard-TeV spectra of TXS 0210+515 and RGB J2042+244. For TXS 0210+515 and 1ES 2037+521, their $\gamma$-ray emissions from $\pi^0$ decay also exceed the one-year sensitivity of WCDA of LHAASO.

In the above, the low-state quasi-simultaneous SEDs of a sample of low TeV luminosity jet-dominated AGNs are reproduced by the one-zone $pp$ model. Constrained by Eq.~(\ref{RS}), relatively small values of $R$ are required in our fitting since the $\gamma$-ray generated in the $\pi^0$ decay is expected to contribute to the TeV spectra. Based on estimated SMBH mass $M_{\rm BH}$ and EBL corrected TeV luminosity $L_{\rm TeV}^{\rm obs}$, most of the values of $R$ are limited to the order of $10^{15}$ cm.  Among them,  $R$ values of M 87 and IC 310 are larger than $10^{16}$ cm since they have the lowest VHE luminosities in our sample. In addition to the low-state emission, fast variabilities (including minute and day scale) are also discovered in the TeV band of some AGNs. Such kind of fast variabilities might arise from a relativistic plasmoid generated in the magnetic reconnection that occurred randomly in the jet \citep{2017ApJ...841...61A}, which may suggest a different origin from the low-state emission, so we do not apply the fast variability timescale to constrain the blob radius in our fitting. In fact, it can be seen from Eq.~(\ref{n_H}) that the $pp$ interactions could be more efficient if a more compact blob is introduced to explain the fast variability. In order to enable $pp$ interactions to make an important contribution to the VHE bands, $\xi$ is set to 1 for most sources, which means that the introduced jet power is basically equivalent to the Eddington luminosity. Only $\xi$ of M 87 is much less than 1, which suggests that M 87 has a large parameter space that can make $pp$ interactions important. On the other hand, the flux of currently detected VHE data points of Mrk 421 and Mrk 501 is much higher than the one-year sensitivity of WCDA of LHAASO, so no matter which process their VHE emission originates from, LHAASO will detect their higher energy TeV emission in the future, unless the TeV spectrum is truncated at the highest energy detected so far. Moreover, our fitting results suggest that the $pp$ interactions can also generate the LHAASO detected TeV emission for IC 310, 3C 264, 1ES 2344+514, RGB J0152+017, TXS 0210+515 and 1ES 2037+521, while other sources, due to the limitation of the currently detected VHE spectra and EBL absorption, cannot be detected by LHAASO. Except for Mrk 421, Mrk 501, RGB J0152+017 and 1ES 1741+196, the emission below 100 GeV and the TeV radiation are decoupled from leptonic and hadronic processes, which propose a possible explanation for detecting different variability patterns between TeV band and other bands \citep{2007ApJ...663..125A, 2012A&A...539L...2A, 2013PASJ...65..109C, 2014ApJ...782...13A, 2016ApJ...819..156B, 2017AA...603A..25A, 2020ApJ...896...41A, 2020MNRAS.492.5354M}. Overall, our results suggest that the $\gamma$-ray emissions of M~87, Mrk 421 and Mrk 501 are most likely to be detected by LHAASO.

\section{Discussion and Conclusion}\label{DC}
In the northern sky, four RGs are discovered as VHE emitters, three of which are studied in this work. Here we will give a brief discussion on the last RG, NGC 1275. At a distance of $\sim$74 Mpc, NGC 1275, also known as 3C 84, 4C+41.07, Perseus A, is the central dominant galaxy in the Perseus cluster. The Very Long Baseline Interferometry measurement infer a range of pc-scale jet angle to the line of sight of $30^\circ-55^\circ$ \citep{1994ApJ...430L..45W, 2006PASJ...58..261A}. The MAGIC collaboration reports its first VHE spectrum with a very steep spectrum extending up to 650 GeV \citep{2012A&A...539L...2A}. In 2016--2017, a VHE flare with a flux doubling timescale of 10 h has been detected by MAGIC telescope \citep{2018A&A...617A..91M}. Ref.~\citep{2014AA...564A...5A} provides two simultaneous SEDs for two campaigns of MAGIC performed between October 2009 and February 2010, and August 2010 and February 2011. The detected VHE spectra are basically in low-states, although VHE light curve shows a hint of variability during the first campaign. From the provided SEDs, we can obtain that the integrated VHE luminosity is about $5\times10^{42}~\rm erg~s^{-1}$. If considering that its SMBH mass is $4\times10^8~M_{\odot}$ \citep{2013MNRAS.429.2315S}, the maximum blob radius $R_{\rm NGC}\approx1\times10^{16}~\rm cm$ can be obtained by Eq.~(\ref{RS}) when applying the one-zone $pp$ model. Let us check the internal $\gamma \gamma$ opacity for the detected highest energy TeV photons. Using the $\delta$-approximation, the energy of soft photons $E_{\rm obs, soft}$ interacting with highest energy TeV photons $E_{\rm obs, TeV}$ can be estimated by \citep{2008ApJ...686..181F}
\begin{equation}
E_{\rm obs, soft} = \frac{2m_{\rm e}c^2\delta_{\rm D}^2}{E_{\rm obs, TeV}} \approx 3.5\rm~eV,
\end{equation}
where $E_{\rm obs, TeV}=600~\rm GeV$ and $\delta_{\rm D}=2$ are taken as suggested in Ref.~\citep{2014AA...564A...5A}. Then the internal $\gamma \gamma$ opacity for the highest energy TeV photons can be estimated by
\begin{equation}
\tau_{\gamma \gamma} = \frac{\sigma_{\gamma \gamma}D_{\rm L}^2\nu F_{\nu}}{R_{\rm NGC}c\delta_{\rm D}^3E_{\rm obs, soft}}\approx 10,
\end{equation}
where $\nu F_{\nu}=3\times10^{-11}~\rm erg~s^{-1}~cm^{-2}$ is the flux of the soft photons and $\sigma_{\gamma \gamma}\approx 1.68\times10^{-25}~\rm cm^2$ is the peak of the $\gamma \gamma$ pair-production cross section. It can be seen that the $pp$ model required relative small blob is opaque to the observed VHE photons. In order to make the blob optically thin to VHE photons, a larger blob radius should be introduced, then $pp$ interactions would not be important for NGC 1275.

In this work, we show that emission from $pp$ interactions can contribute to the VHE emissions of AGNs. Here we discuss some other possibilities. For the emission around 1 TeV, the one-zone leptonic model is likely to give a reasonable explanation \citep{2012ApJ...752..157Z, 2017MNRAS.464..599D}, except for the hard-TeV spectrum, since the KN effect will soften the spectrum naturally, unless introducing an extremely high Doppler factor \citep{2009MNRAS.399L..59T, 2012A&A...544A.142A, 2014ApJ...782...13A}. The proton synchrotron emission can explain the TeV emission as well if assuming a strong magnetic field (10--100 G) \citep{2013ApJ...768...54B}. The one-zone $p\gamma$ model faces difficulties in explaining $\sim$1 TeV radiation. If considering the condition where cross section of photopion interaction peaks due to the $\bigtriangleup^+(1232)$ resonance is \citep{1997PhRvL..78.2292W}
\begin{equation}
E_{\rm p}E_{\rm soft} \simeq 0.3~\rm GeV^2,
\end{equation}
where $E_{\rm p}$ and $E_{\rm soft}$ are the proton and soft photon energies in the comoving frame, respectively, the required soft photon energy is about 1 MeV in the observers' frame. While the number density of MeV photons in jet-dominated AGNs is usually quite low, so an extremely large jet power have to be introduced \citep{2014ApJ...783..108C, 2022ApJ...925L..19C}. Some studies reduce the jet power by assuming a very large minimum proton Lorentz factor \citep{2016ApJ...830...81F, 2017APh....89...14F} or by assuming the existence of an extremely dense MeV photon field that has not yet been discovered (current MeV detectors are relatively insensitive) \citep{2013PhRvD..87j3015S, 2015EPJC...75..273S, 2018EPJC...78..484S, 2019ApJ...884L..17S, 2020ApJ...898..103S}. If protons with energy $> 10^{19}~\rm eV$ can be accelerated and escape the blazar jet, the cascade emission generated in the intergalactic space through $p\gamma$ interactions by interacting with EBL and CMB may also contribute to TeV emission without significant variability \citep{2011APh....35..135E, 2012ApJ...757..183P, 2020ApJ...889..149D, 2022A&A...658L...6D}. For the emission around 10 TeV, due to the KN effect, it is difficult for the one-zone leptonic model to interpret it in a reasonable parameter space. If the one-zone $p\gamma$ model is applied, the same problem when applying it to explain $\sim$1 TeV emission arise again. An extremely large jet power needs to be introduced, because the target photons are in the hard X-ray band with a very low number density. If the emission around 10 TeV will be detected in the near future, the emitting region producing the 10 TeV emission is likely to be decoupled from the blob generating the typical SED \citep{2021ApJ...906...51X, 2022PhRvD.105b3005W}. 

To summarize, we propose a one-zone $pp$ model to revisit the quasi-simultaneous multi-wavelength SEDs of a sample of low TeV luminosity AGNs in this work. Contrary to conventional views, our numerical modeling results suggest that $pp$ interactions in the jet are important under certain conditions and generate VHE $\gamma$-ray. In the modeling, the emission with energy below $\sim$ 1 TeV is explained by conventional synchrotron and SSC emissions from primary relativistic electrons. From our results, we suggest that M~87, Mrk 421 and Mrk 501 are the most likely AGNs to be detected by LHAASO. For M 87, due to its extremely low TeV luminosity, there is a large parameter space for $pp$ interactions to generate detectable VHE $\gamma$-ray. For Mrk 421 and Mrk 501, their IC emissions extend to higher energy band that could be detected by LHAASO naturally. It can be seen that the observed VHE spectra of Mrk 421 and Mrk 501 do not show truncated feature and the flux of VHE data points is much higher than the one-year LHAASO sensitivity, so no matter which process their VHE emission originates from, LHAASO will detect their higher energy TeV emission in the future.

\section*{Acknowledgements}
We thank the anonymous referees for insightful comments and constructive suggestions. This work is supported by the National Natural Science Foundation of China (NSFC) under the grants No. 12203043 and 12203024.

\appendix
\section{The internal $\gamma \gamma$ opacity}\label{gg}
In this work, our main purpose is to show that $pp$ interactions can generate detectable VHE emission. As indicated by Eq.~(\ref{RS}), a relative compact blob is required in the one-zone $pp$ model, therefore one may curious if the VHE photons would be absorbed due to the internal $\gamma \gamma$ annihilation. Here we calculate the internal $\gamma \gamma$ opacity $\tau_{\gamma \gamma}$ for each object in our sample. For uniform isotropic photon fields \citep{2009herb.book.....D}, $\tau_{\gamma \gamma}$ can be calculated by
\begin{eqnarray}
\tau_{\gamma \gamma}(\epsilon_1) = \frac{R \pi r_{\rm e}^2}{\epsilon_1^2}
\int^\infty_{1/\epsilon_1} d\epsilon\ n_{\rm soft}(\epsilon)\ \bar{\phi}(s_0)\epsilon^{-2},\ 
\end{eqnarray}
where $\epsilon$ and $\epsilon_1$ are the dimensionless energies of low-energy and high-energy photons, $n_{\rm soft}$ is the number density of soft photons, $s_0=\epsilon \epsilon_1$, 
\begin{eqnarray}
\bar{\phi}(s_0) = \frac{1+\beta_0^2}{1-\beta_0^2}\ln w_0 - 
\beta_0^2\ln w_0 - \frac{4\beta_0}{1 - \beta_0^2}
\\ \nonumber 
+ 2\beta_0 + 4\ln w_0 \ln(1+w_0) - 4L(w_0)\ ,
\end{eqnarray}
$\beta_0^2 = 1 - 1/s_0$, $w_0=(1+\beta_0)/(1-\beta_0)$, and 
\begin{eqnarray}
L(w_0) = \int^{w_0}_1 dw\ w^{-1}\ln(1+w)\ .
\end{eqnarray}

As mentioned before, since AGNs in our sample are FR I RGs and HBLs, photons from external fields are neglected in our modeling. Therefore, curves of $\tau_{\gamma \gamma}$ of AGNs shown in Fig.~\ref{opa} are contributed by the synchrotron photons from primary electrons. It can be seen that VHE photons can escape, since the internal $\gamma \gamma$ opacity is smaller than unity.

{On the other hand, the broad-line region (BLR) luminosities $L_{\rm BLR}\approx 2.3\times10^{41}~\rm erg~s^{-1}$ of Mrk 421 and Mrk 501 are given in Ref.~\citep{2012ApJ...759..114C, 2014MNRAS.441.3375X, 2016MNRAS.463.3038X}, which may suggest the existence of weak external photon fields. Here, we estimate the energy densities of two typical external photon fields, which are BLR and the dusty torus (DT), as a function of the distance along the jet, $r$, in the comoving frame by \citep{2012ApJ...754..114H}, 
\begin{equation}
u_{\rm BLR} = \frac{\Gamma^2L_{\rm BLR}}{4\pi r_{\rm BLR}^2c[1+(r/r_{\rm BLR})^3]}
\end{equation}
and
\begin{equation}
u_{\rm DT} = \frac{\Gamma^2L_{\rm DT}}{4\pi r_{\rm DT}^2c[1+(r/r_{\rm DT})^4]},
\end{equation}
where the value of $L_{\rm DT}$ is assumed to be the same of $L_{\rm BLR}$, $r_{\rm BLR} = 0.1(L_{\rm BLR}/10^{45}\rm erg s^{-1})^{1/2}$pc and $r_{\rm DT} = 2.5(L_{\rm DT}/10^{45}\rm erg s^{-1})^{1/2}$pc are the characteristic radius of the BLR in the AGN frame. The radiation from both the BLR and DT radiation is taken as an isotropic graybody with a peak at $2\times10^{15}\Gamma$ Hz \citep{2008MNRAS.386..945T} and $3\times10^{13}\Gamma$ Hz \citep{2007ApJ...660..117C} in the jet comoving frame, respectively. From the above equations, it can be found that the energy densities of external photon fields highly depend on the blob's position in the jet. Under the assumption that the jet has a conical structure and the blob occupy the entire cross-section of the jet, we further check if photons from BLR and DT would absorb VHE photons in the blob. By adopting the jet half opening angles $\alpha$ of Mrk 421 and Mrk 501 are both $2.5^\circ$ \citep{2019ApJ...870...28F}, we speculate that the blobs of Mrk 421 and Mrk 501 are at the distances $r=R/tan~\alpha$ of $4\times10^{-2}$ pc and $6\times10^{-3}$ pc, respectively, which are outside the BLR and within the DT. In left panels of Fig.~\ref{opa3}, we show the $\tau_{\gamma \gamma}$ including the photons from primary electrons, BLR and DT. It can be seen that $\tau_{\gamma \gamma}$ is lower than unity, which suggests that VHE photons would not be absorbed. Please note that the $\tau_{\gamma \gamma}$ contributed by DT is only slightly lower than unity. This tension can be alleviated by introducing $L_{\rm DT}<L_{\rm BLR}$ or assuming the blob does not occupy the entire cross-section of the jet so that it can be placed at a larger distance. In right panels of Fig.~\ref{opa3}, we show $\tau_{\gamma \gamma}$ at four energies as a function of the blob distance $r$. It can be seen that with the decrease of $r$, i.e., the blob position is gradually closer to the SMBH, the energy density of the BLR will increase significantly, thereby increasing the $\tau_{\gamma \gamma}$ at 0.1-1 TeV energy range. This result suggests that the blobs should not be too close to the SMBH, otherwise even a weak external photon field may still absorb VHE photons. Please note that the further consideration of BLR and DT may make EC cooling non-negligible, but similar fitting results can still be obtained by fine-tuning free parameters.

\begin{figure*}
\centering
\subfloat{
\includegraphics[width=1\columnwidth]{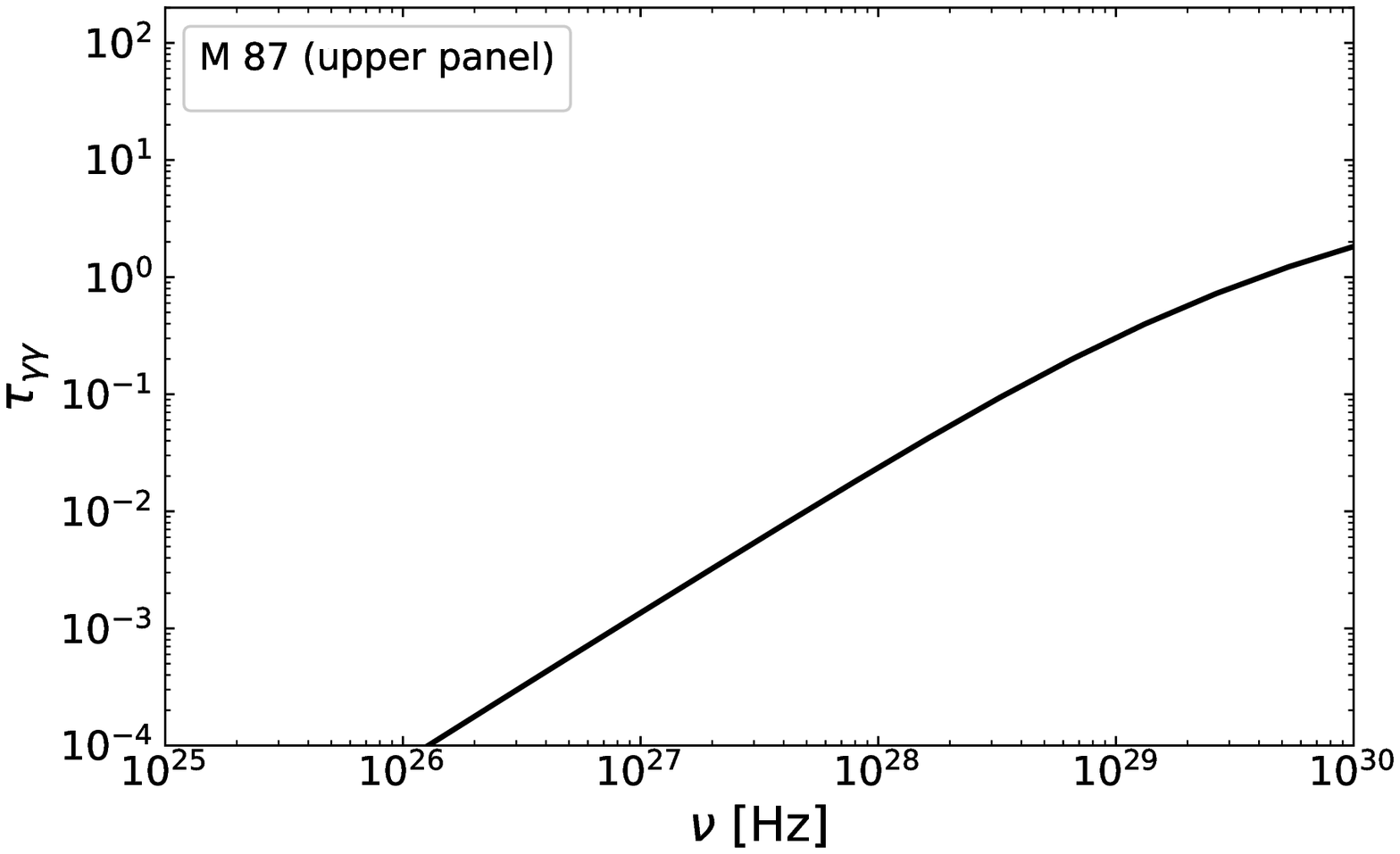}
}\hspace{-5mm}
\quad
\subfloat{
\includegraphics[width=1\columnwidth]{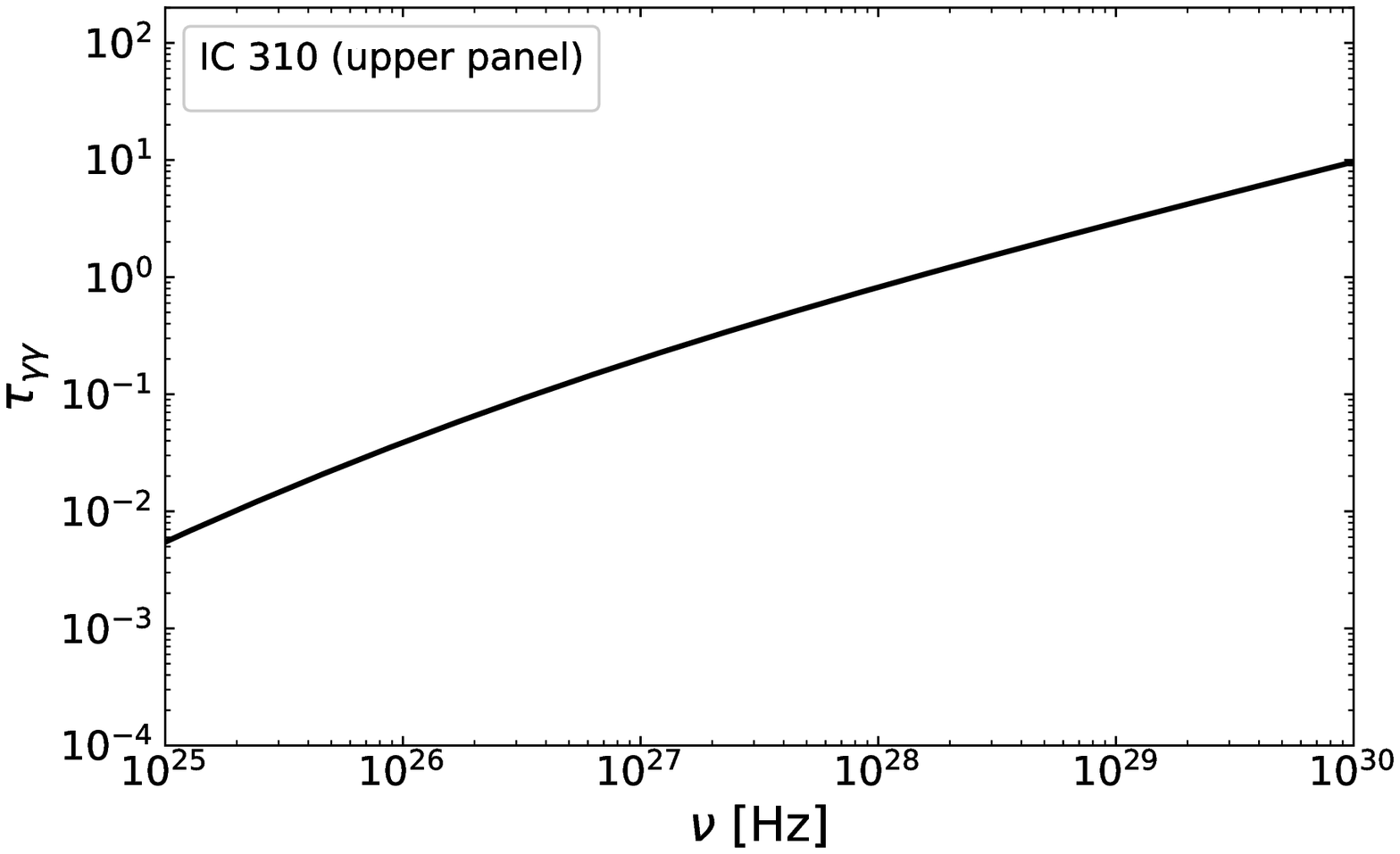}
}\hspace{-5mm}
\subfloat{
\includegraphics[width=1\columnwidth]{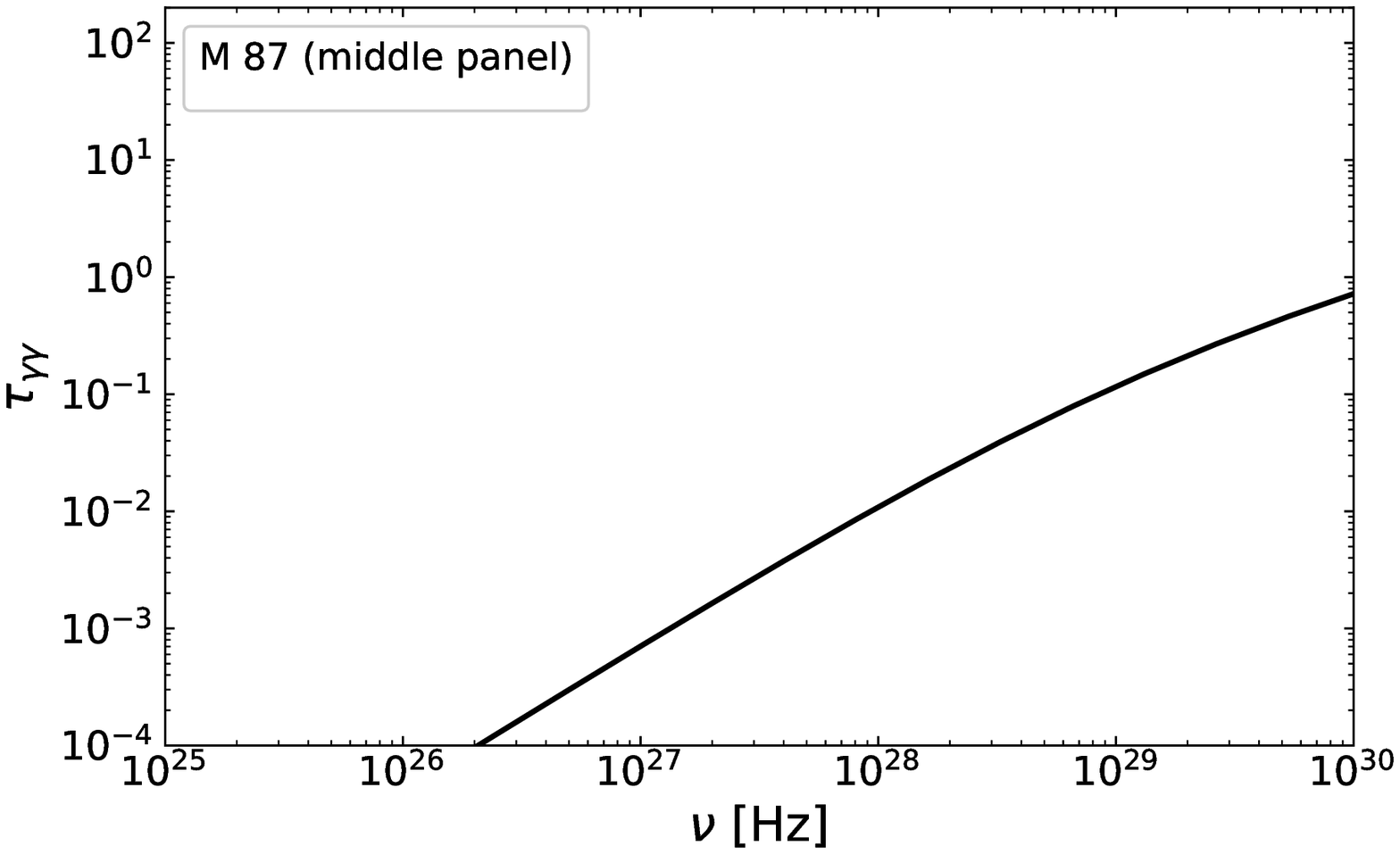}
}\hspace{-5mm}
\quad
\subfloat{
\includegraphics[width=1\columnwidth]{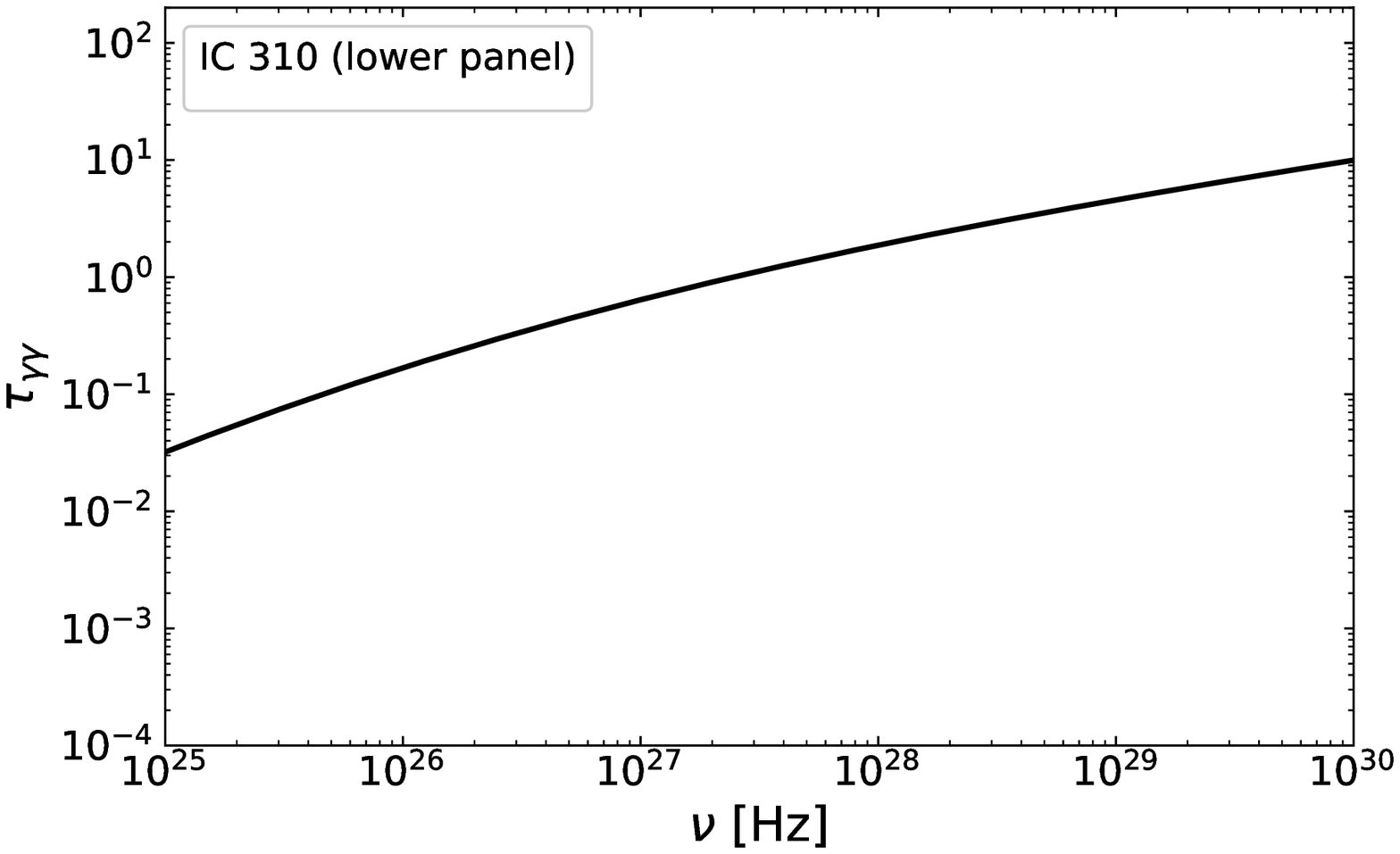}
}\hspace{-5mm}
\quad
\subfloat{
\includegraphics[width=1\columnwidth]{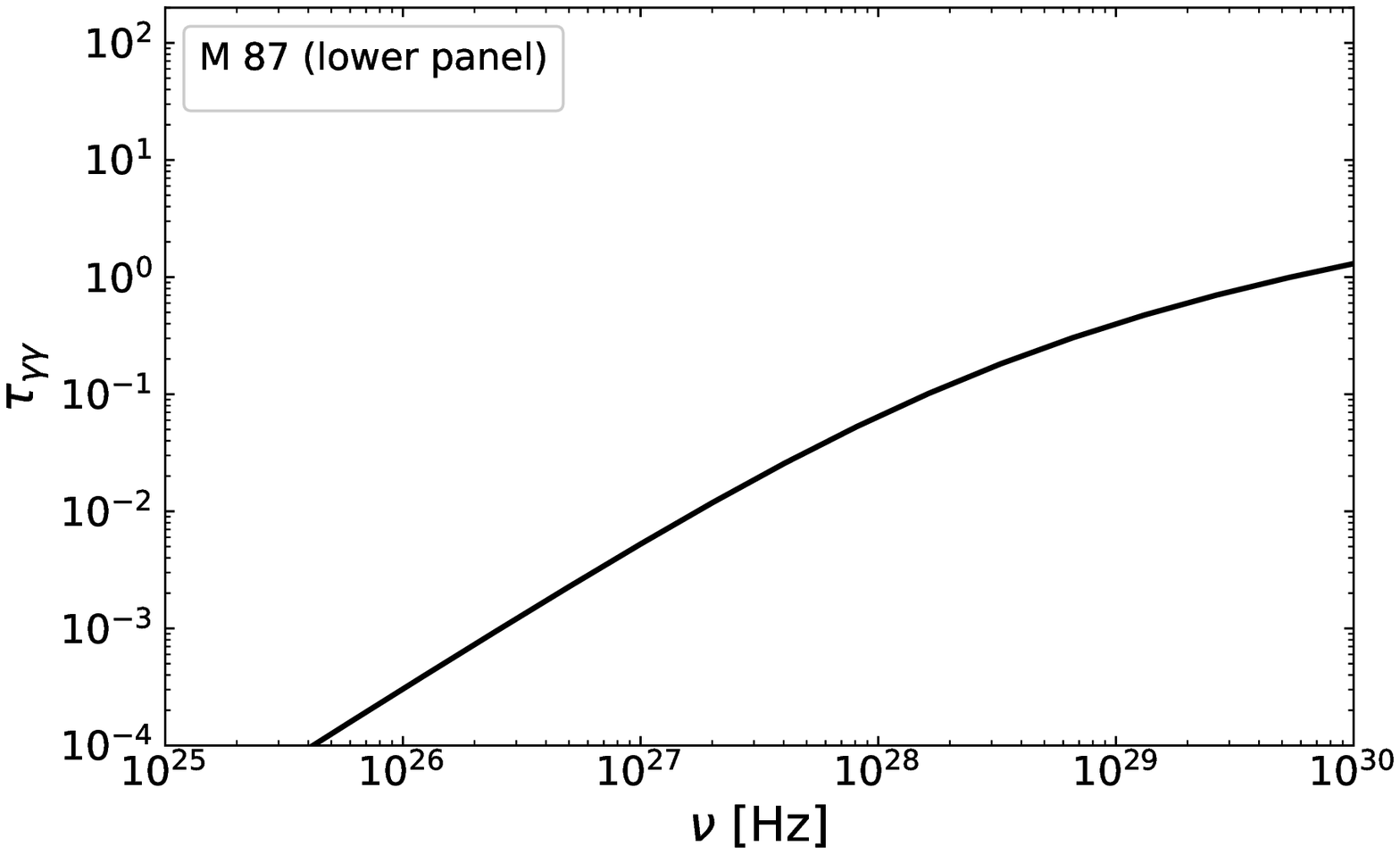}
}\hspace{-5mm}
\quad
\subfloat{
\includegraphics[width=1\columnwidth]{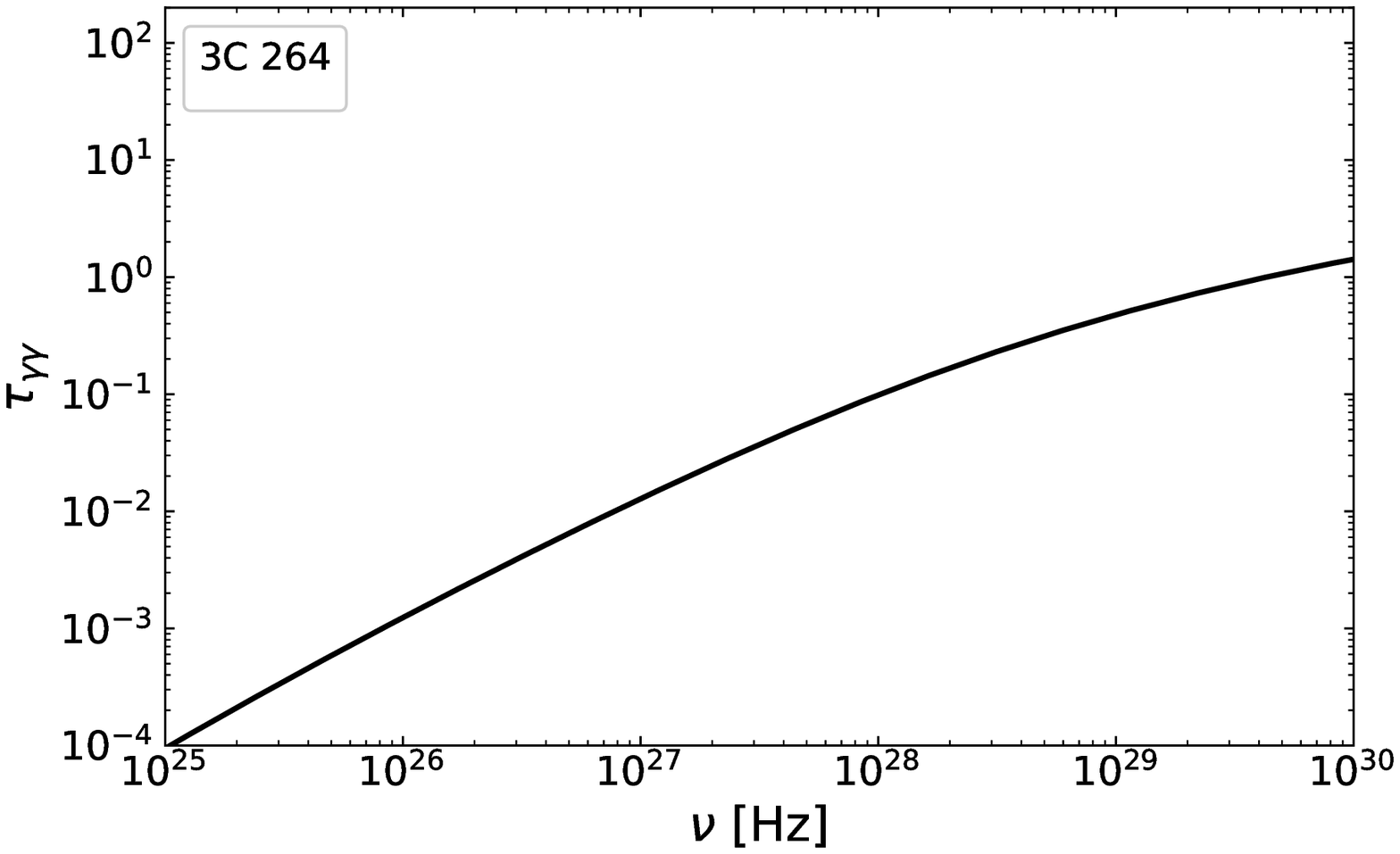}
}\hspace{-5mm}
\quad
\subfloat{
\includegraphics[width=1\columnwidth]{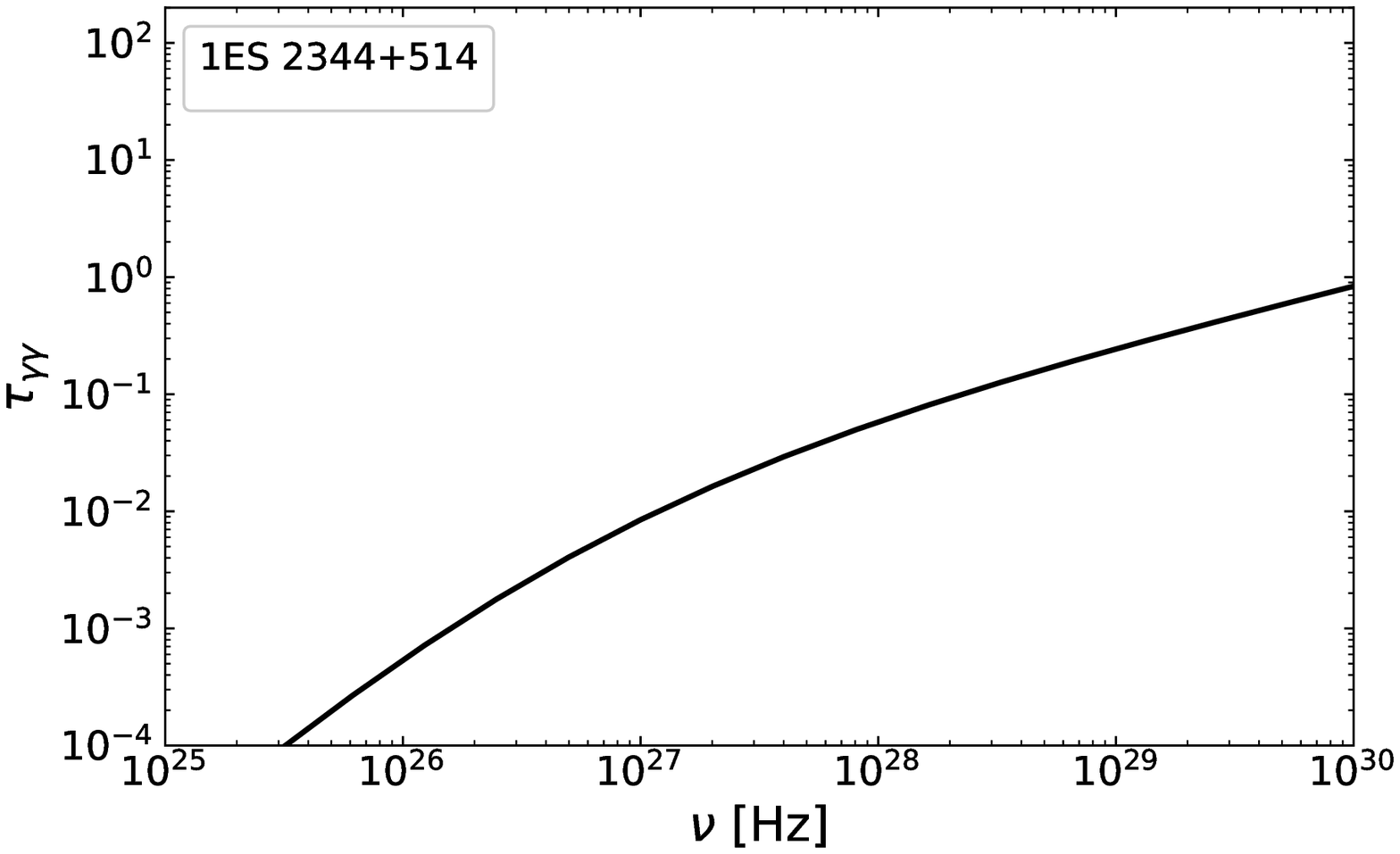}
}\hspace{-5mm}
\quad
\subfloat{
\includegraphics[width=1\columnwidth]{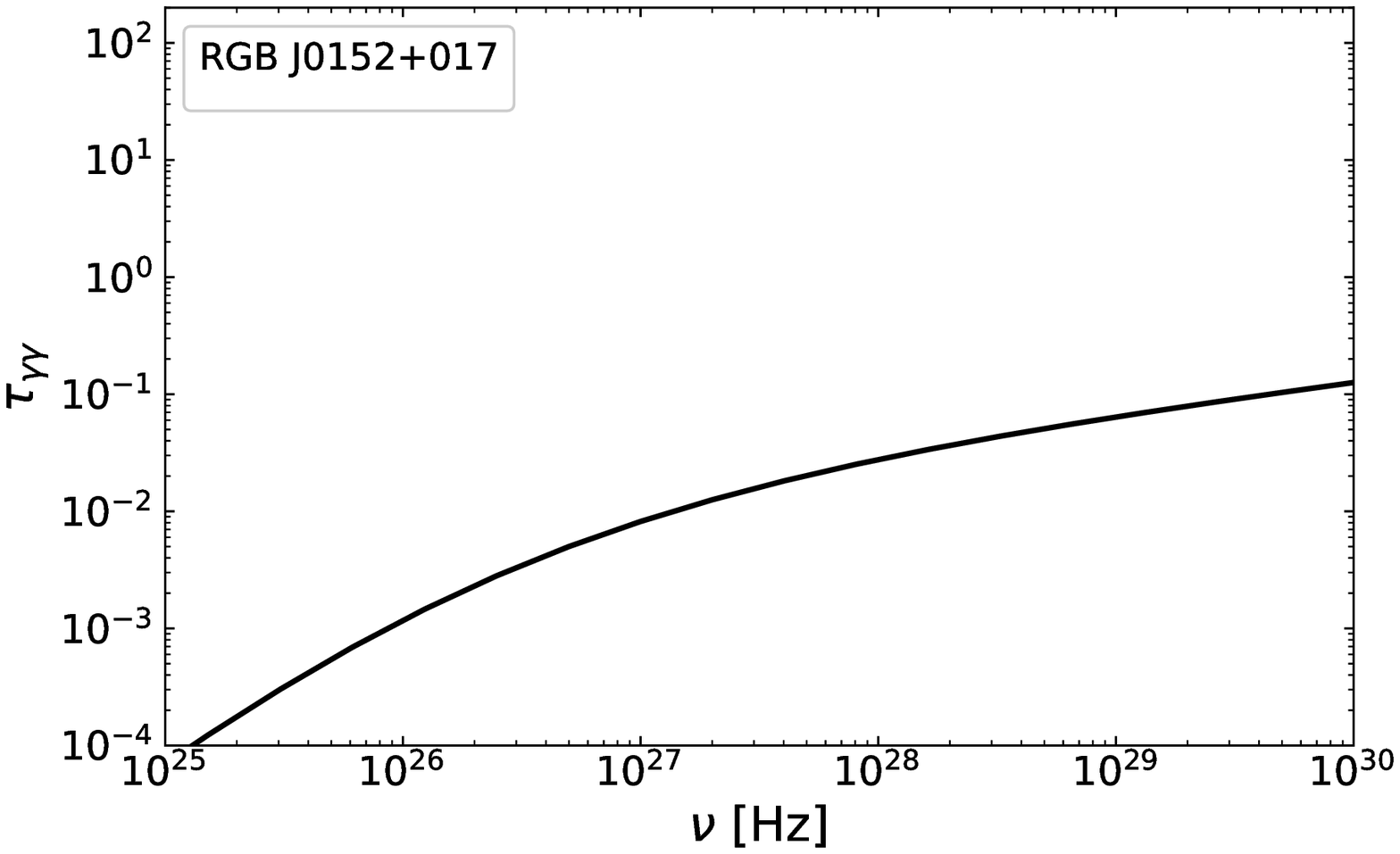}
}
\caption{The internal $\gamma \gamma$ opacity $\tau_{\gamma \gamma}$ as a function of the photon frequency in the observers' frame for M~87, IC~310, 3C 264, 1ES 2344+514, RGB J0152+017, respectively.
}
\end{figure*}

\addtocounter{figure}{-1}
\begin{figure*}
\centering
\subfloat{
\includegraphics[width=1\columnwidth]{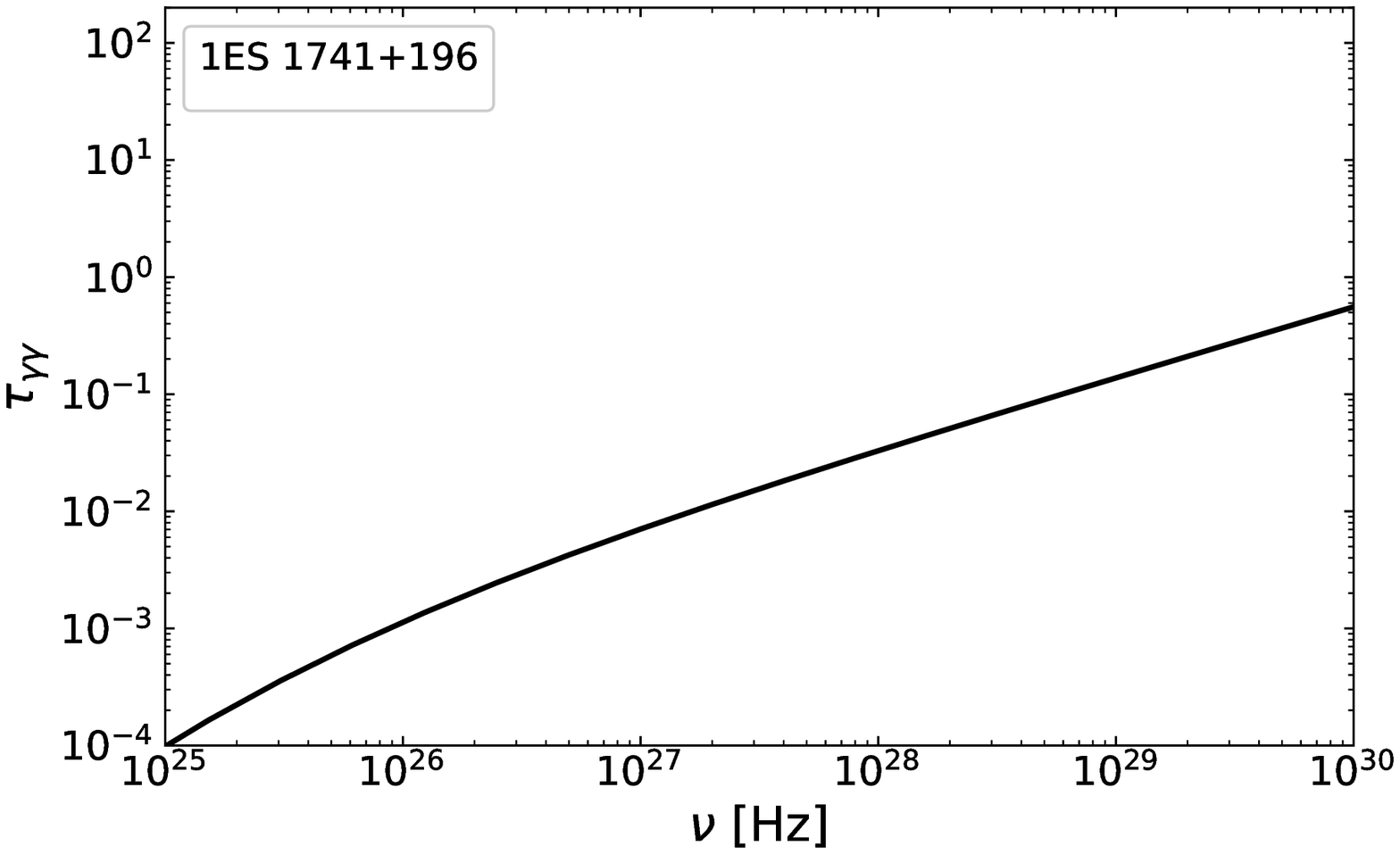}
}\hspace{-5mm}
\quad
\subfloat{
\includegraphics[width=1\columnwidth]{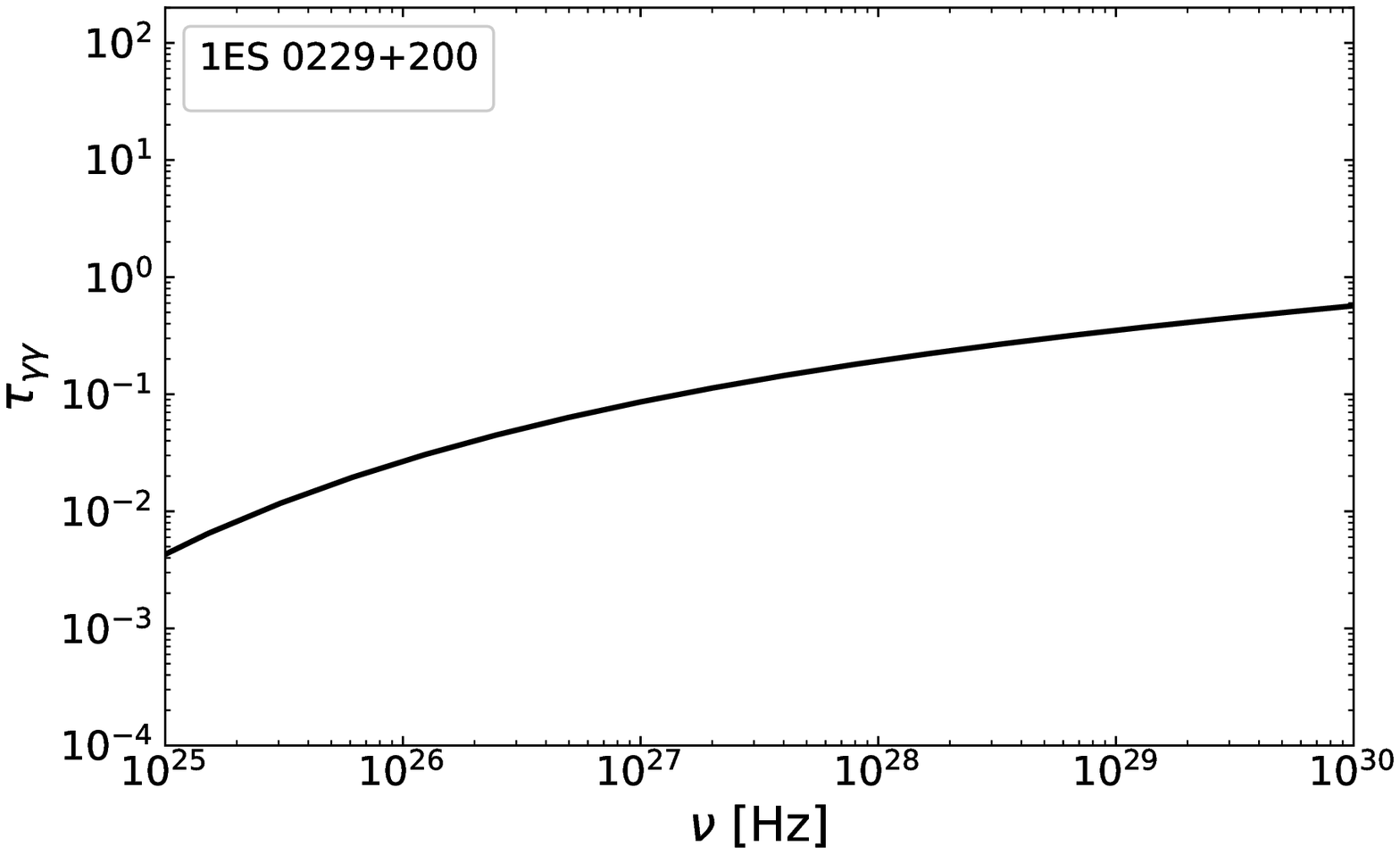}
}\hspace{-5mm}
\quad
\subfloat{
\includegraphics[width=1\columnwidth]{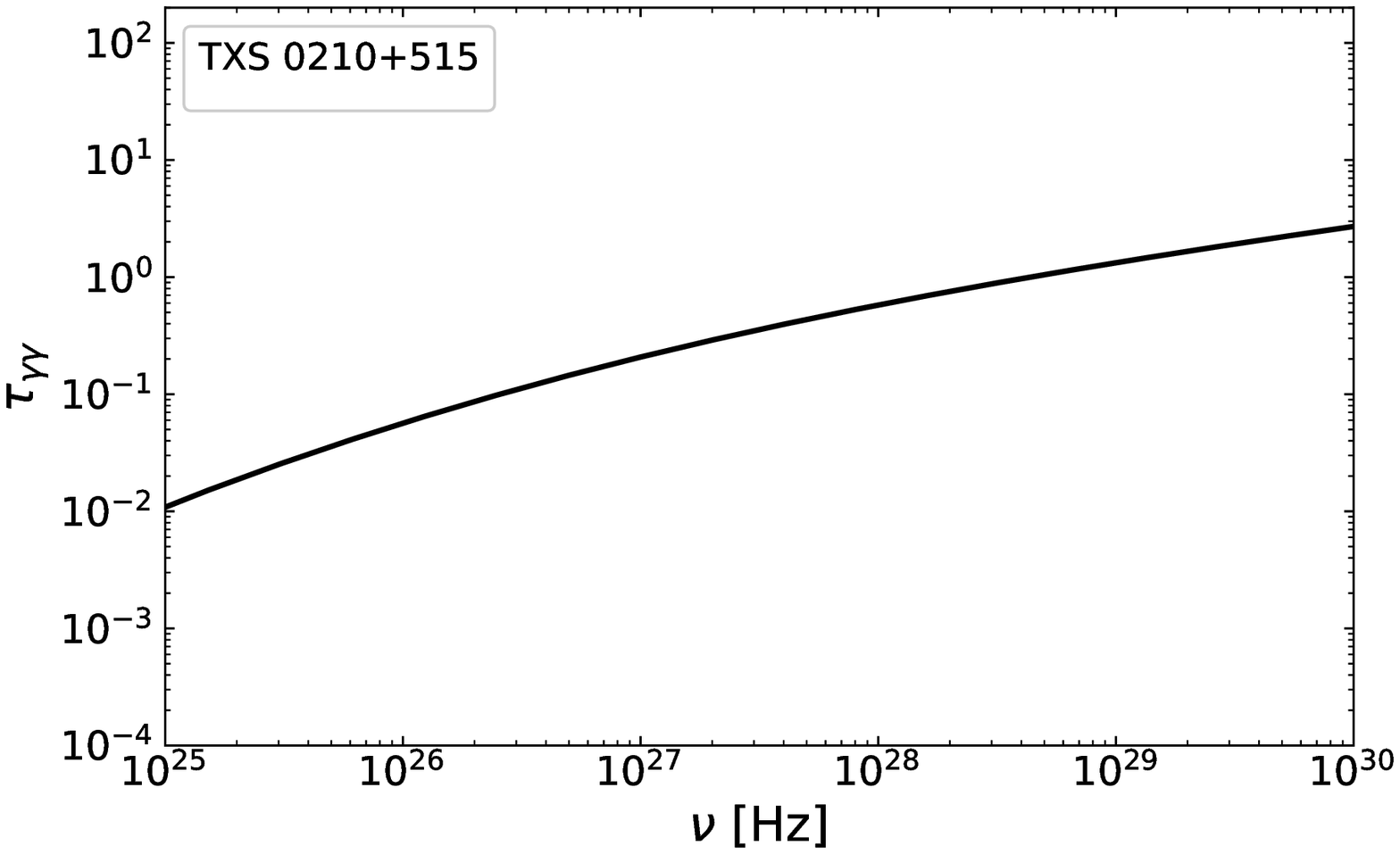}
}\hspace{-5mm}
\quad
\subfloat{
\includegraphics[width=1\columnwidth]{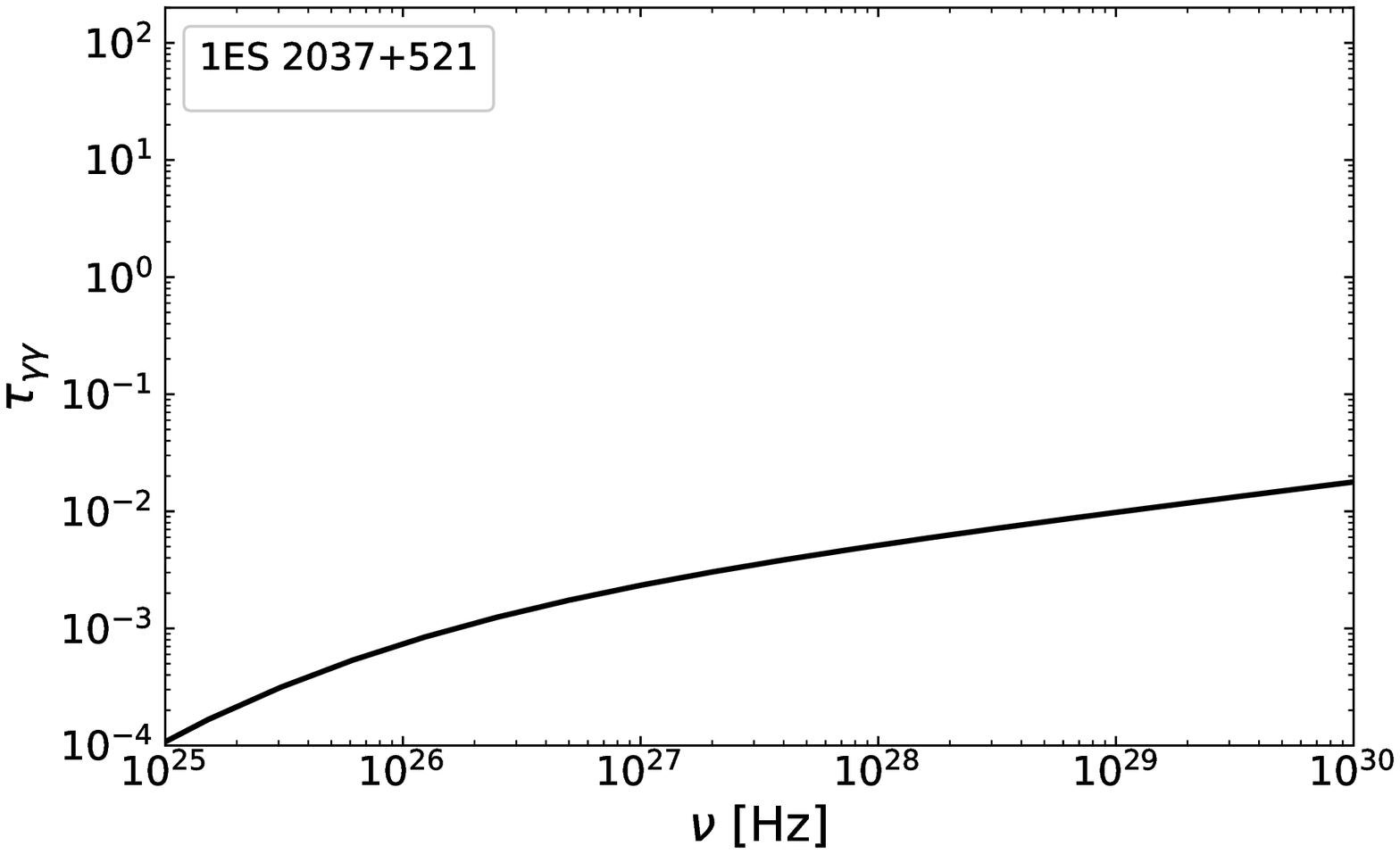}
}\hspace{-5mm}
\quad
\subfloat{
\includegraphics[width=1\columnwidth]{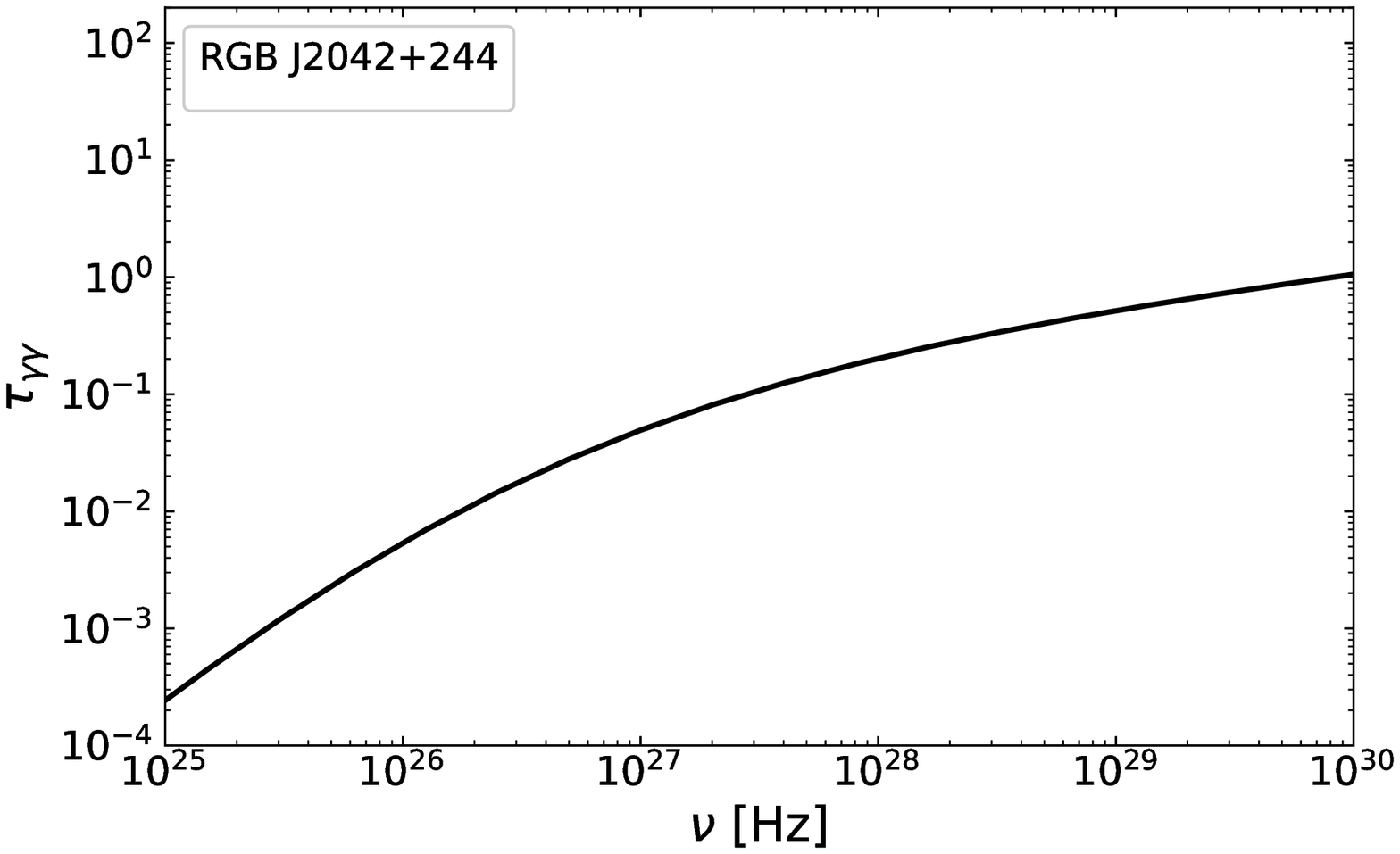}
}
\caption{The internal $\gamma \gamma$ opacity $\tau_{\gamma \gamma}$ as a function of the photon frequency in the observers' frame for 1ES 1741+196, 1ES 0229+200, TXS 0210+515, 1ES 2037+521, RGB J2042+244, respectively.
\label{opa}}
\end{figure*}

\begin{figure*}
\centering
\subfloat{
\includegraphics[width=1\columnwidth]{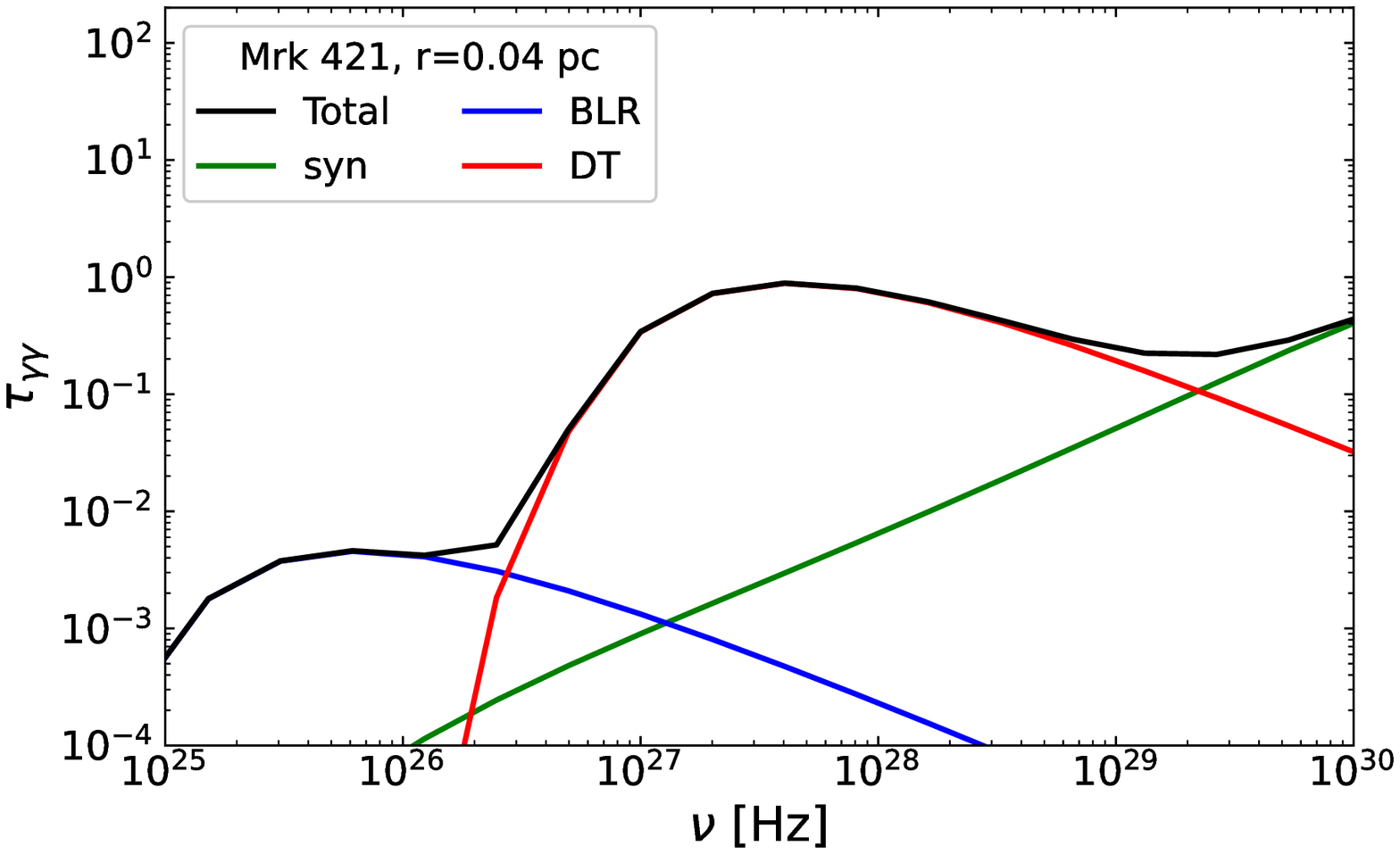}
}\hspace{-5mm}
\quad
\subfloat{
\includegraphics[width=1\columnwidth]{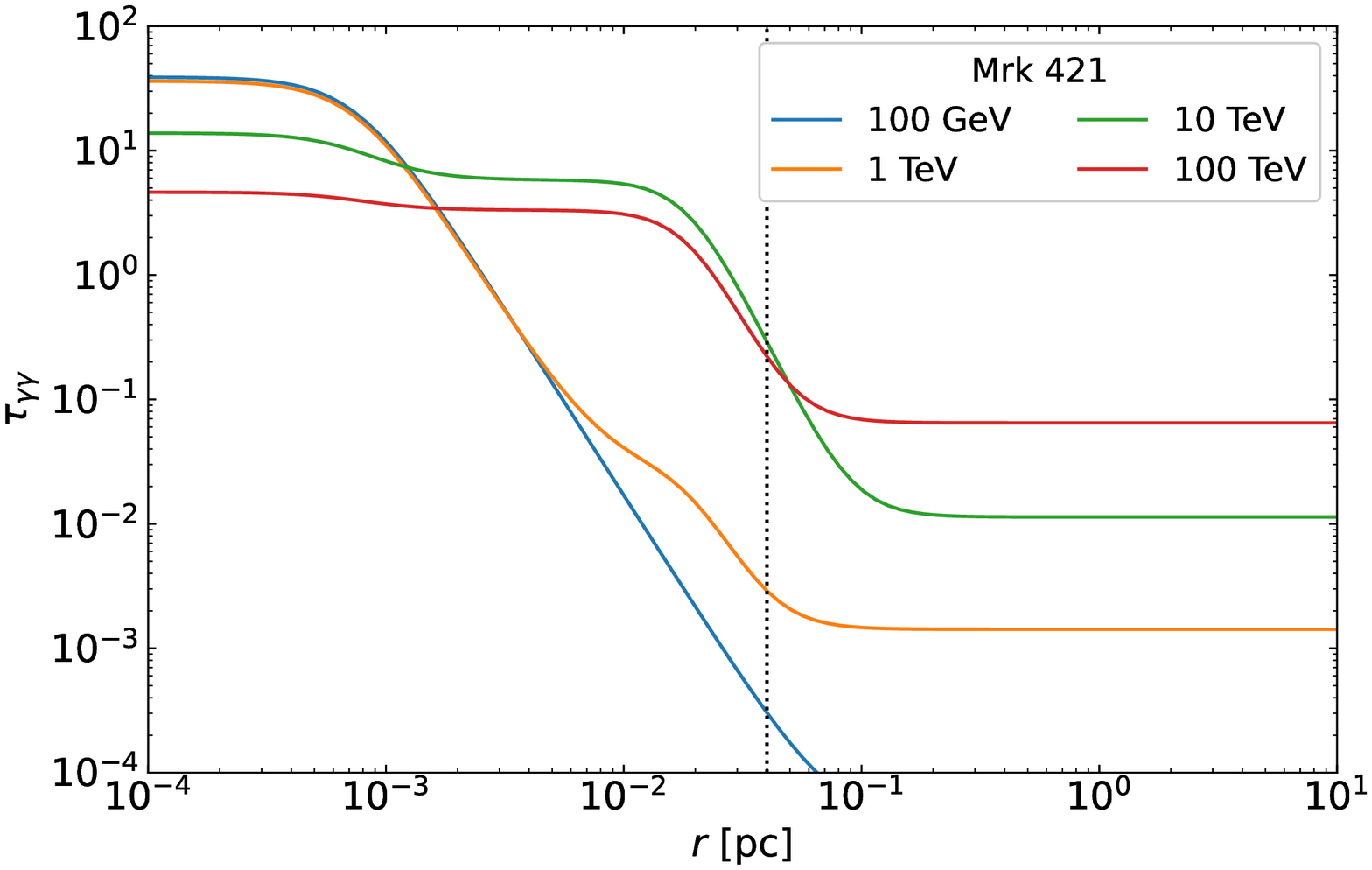}
}\hspace{-5mm}
\quad
\subfloat{
\includegraphics[width=1\columnwidth]{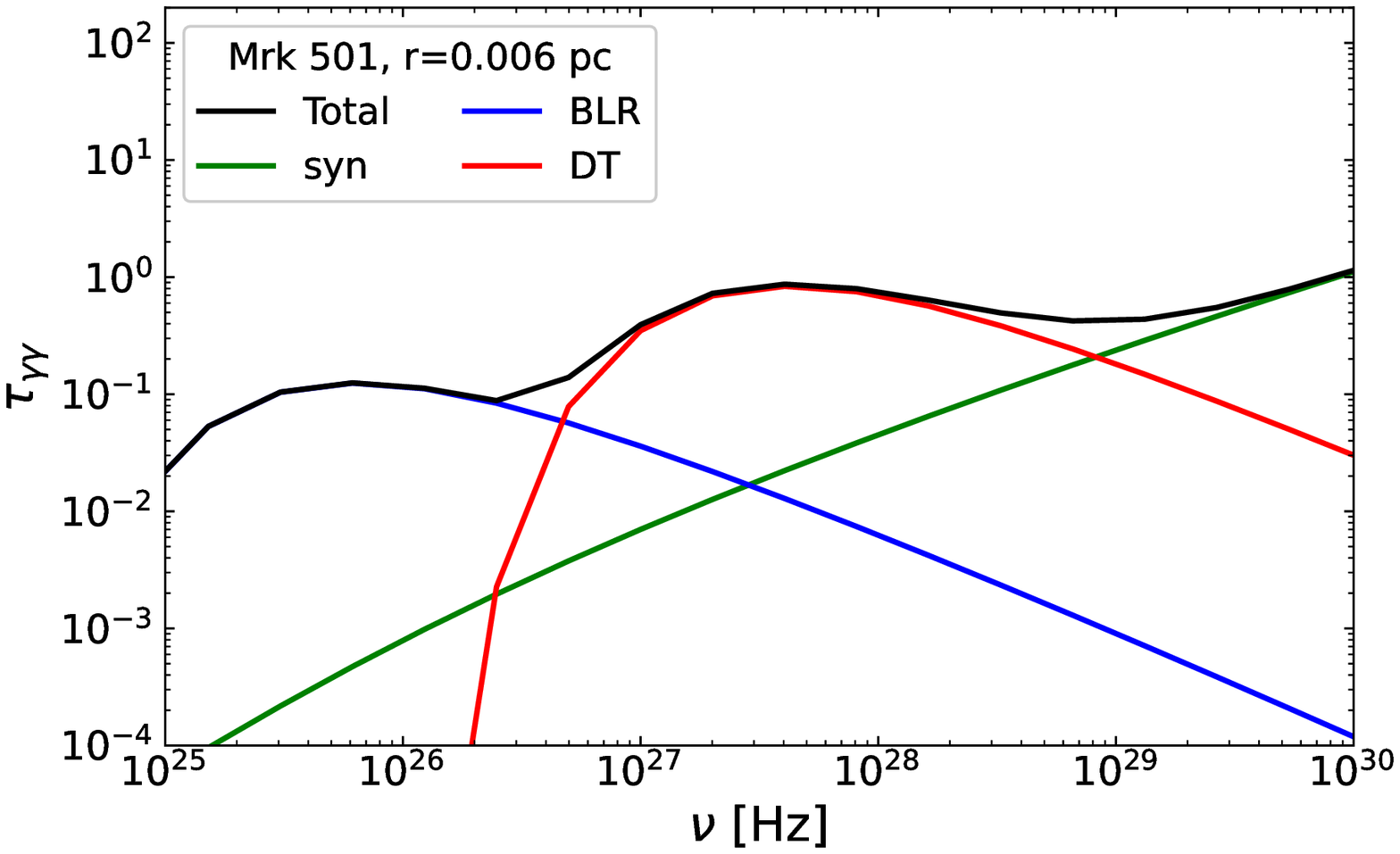}
}\hspace{-5mm}
\quad
\subfloat{
\includegraphics[width=1\columnwidth]{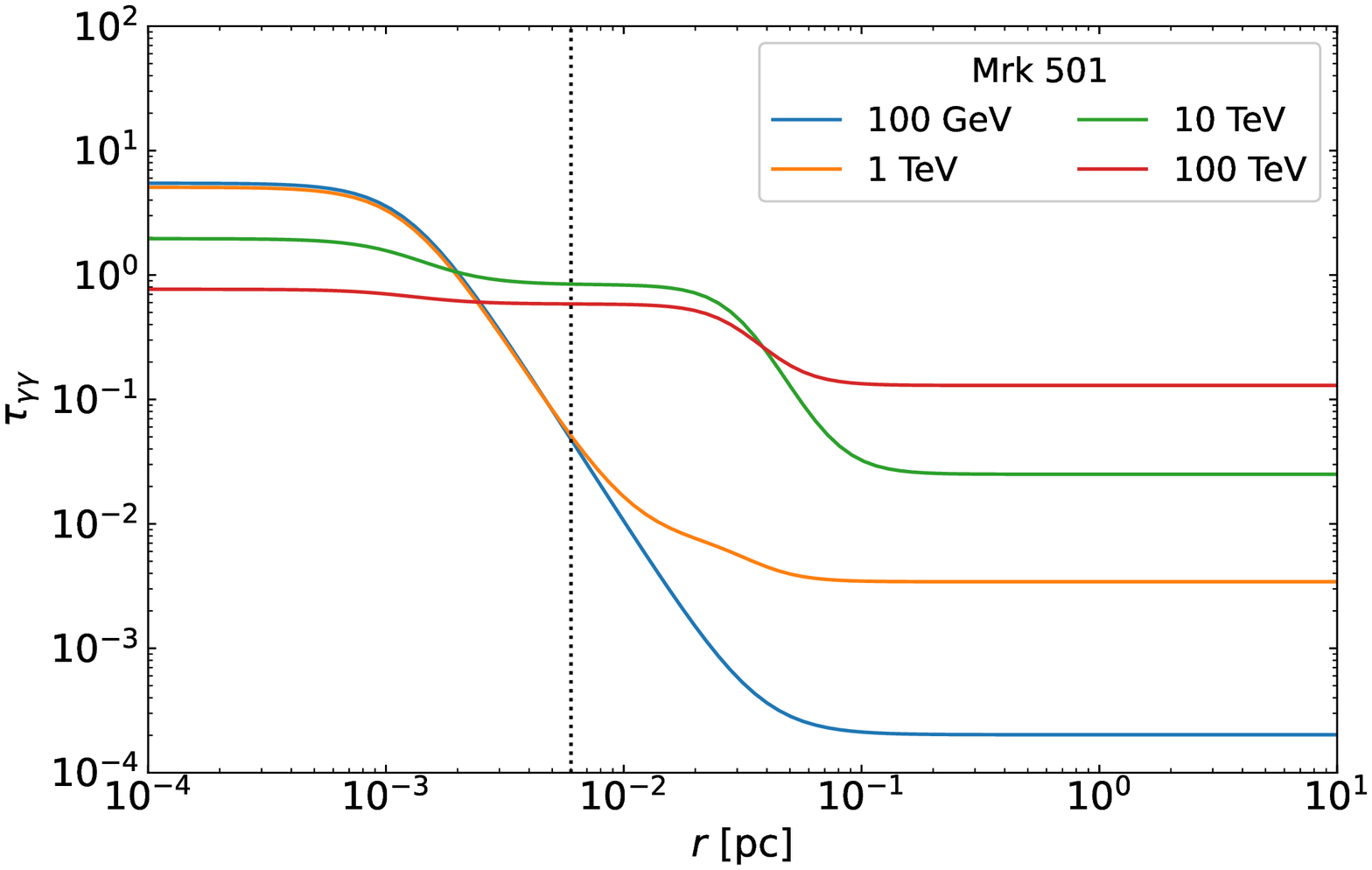}
}
\caption{Left panels: The internal $\gamma \gamma$ opacity $\tau_{\gamma \gamma}$ as a function of the photon frequency in the observers' frame for Mrk 421 and Mrk 501, respectively. Right panels: The internal $\gamma \gamma$ opacity $\tau_{\gamma \gamma}$ as a function of distance from the SMBH in the observers' frame for Mrk 421 and Mrk 501, respectively. The meaning of all curves is explained in the inset legends. 
\label{opa3}}
\end{figure*}

\bibliographystyle{apsrev}
\bibliography{ms.bib}

\end{document}